\documentclass[twocolumn]{aastex7}
\usepackage{xspace}
\usepackage{multirow}
\usepackage{natbib}
\usepackage{mathtools}
\usepackage{comment}
\usepackage{rotating}

\newcommand{\be}{\begin{equation}}
\newcommand{\ee}{\end{equation}}
\newcommand{\bea}{\begin{eqnarray}}
\newcommand{\eea}{\end{eqnarray}}

\newcommand{\jwst}{\textit{JWST}\xspace}

\defcitealias{harris_reinacampos2023} {Paper I}
\defcitealias{harris_reinacampos2024} {Paper II}

\nolinenumbers

\begin{document}

\title{\textit{JWST} Photometry of Globular Cluster Populations in MACS0417.5-1154}

\author[orcid=0000-0001-8762-5772,sname='Harris']{William E. Harris} 
\affiliation{Department of Physics \& Astronomy, McMaster University, Hamilton ON L8S 4M1, Canada}
\email[show]{harrisw@mcmaster.ca}

\author[orcid=0000-0002-8556-4280,sname='Reina-Campos']{Marta Reina-Campos}
\affiliation{Canadian Institute for Theoretical Astrophysics (CITA), University of Toronto, 60 St George St, Toronto, M5S 3H8, Canada}
\affiliation{Department of Physics \& Astronomy, McMaster University, Hamilton ON L8S 4M1, Canada}
\email[show]{reinacampos@cita.utoronto.ca}

\author[0000-0001-7599-1967,sname='Kaitlyn Keatley']{Kaitlyn E. Keatley}
\affiliation{Department of Physics \& Astronomy, McMaster University,  Hamilton ON L8S 4M1, Canada}
\email{keatleyk@mcmaster.ca}

\author[0000-0001-5984-0395]{Maru\u{s}a Brada\u{c}}
\affiliation{University of Ljubljana, Faculty of Mathematics and Physics, Jadranska ulica 19, SI-1000 Ljubljana, Slovenia}
\email{marusa.bradac@fmf.uni-lj.si}

\author[0000-0003-3243-9969]{Nicholas S. Martis}
\affiliation{University of Ljubljana, Faculty of Mathematics and Physics, Jadranska ulica 19, SI-1000 Ljubljana, Slovenia}
\email{nicholas.martis@fmf.uni-lj.si}

\author[0000-0002-9330-9108]{Adam Muzzin}
\affiliation{Department of Physics and Astronomy, York University, 4700 Keele St. Toronto, Ontario, M3J 1P3, Canada}
\email{muzzin@yorku.ca} 

\author[]{Ga\"el Noirot}
\affiliation{Space Telescope Science Institute, 3700 San Martin Drive, Baltimore, Maryland 21218, USA}
\email{gnoirot@stsci.edu}

\author[0000-0001-8830-2166]{Ghassan T. E. Sarrouh}
\affiliation{Department of Physics and Astronomy, York University, 4700 Keele St. Toronto, Ontario, M3J 1P3, Canada}
\email{gsarrouh@yorku.ca}

\author[0000-0002-7712-7857]{Marcin Sawicki}
\affiliation{Department of Astronomy and Physics and Institute for Computational Astrophysics, Saint Mary’s University, 923 Robie Street, Halifax, Nova Scotia B3H 3C3, Canada}
\email{marcin.sawicki@smu.ca}

\author[0000-0002-4201-7367]{Chris J. Willott}
\affiliation{National Research Council of Canada, Herzberg Astronomy \& Astrophysics Research Centre, 5071 West Saanich Road, Victoria,BC V9E 2E7, Canada}
\email{christopher.willott@nrc-cnrc.gc.ca}

\author[0000-0001-7549-5560]{Samantha C. Berek}
\affiliation{David A. Dunlap Department of Astronomy \& Astrophysics, University of Toronto, 50 St George Street, Toronto, ON M5S 3H4, Canada}
\affiliation{Dunlap Institute for Astronomy \& Astrophysics, University of Toronto, 50 St George Street, Toronto, ON M5S 3H4, Canada}
\affiliation{Data Sciences Institute, University of Toronto, 17th Floor, Ontario Power Building, 700 University Ave, Toronto, ON M5G 1Z5, Canada}
\email{sam.berek@mail.utoronto.ca}

\correspondingauthor{WEH}
\email{harrisw@mcmaster.ca}

\begin{abstract}
Deep \jwst imaging of the massive galaxy cluster MACS0417.5-1154, at redshift $z=0.443$, reveals a huge population of globular clusters (GCs) and Ultra-Compact Dwarfs (UCDs)  primarily distributed around its single central giant galaxy (BCG). We present NIRCam/SWC photometry of the GC system in four bands (F090W, F115W, F150W, F200W). The spatial distribution of the system matches well in radial and ellipticity profile with the high elongation ($b/a \simeq 0.5$) of the BCG halo light. The total GC population within MACS0417 is estimated to be near $1.5 \times 10^5$, similar to the systems in Abell 2744, Coma, and other galaxy clusters with comparable masses. With similar results for GC photometry in hand from other lensing clusters at a range of redshifts, it is now possible to trace on purely observational grounds the luminosity evolution of GC systems over many Gigayears of lookback time, as seen through their color-magnitude diagrams. We show this sequence for five systems reaching to lookback times of more than 7 Gyr. A systematic change in the GC/UCD sequence with lookback time is clearly visible, near what is expected for age-fading of a simple stellar population with time.  Lastly, we evaluate the effectiveness of the various JWST NIRCam filters for broadband photometry of GC systems as a function of redshift, as an aid to planning further studies.

\end{abstract}


\section{Introduction}

The James Webb Space Telescope (\jwst) has opened up a huge new range of cosmological space for a wide array of observational studies. Among these is the ability to measure entire populations of globular clusters (GCs) out to redshifts as distant as $z \sim 0.5$ and beyond, and thus stages of their evolution many Gyr earlier than the present zero-redshift epoch.
In \citet{harris_reinacampos2023,harris_reinacampos2024} (hereafter Papers I and II), we began a series of studies of GC populations in rich, distant clusters of galaxies with deep photometry. The first target was the central region of Abell 2744 at $z=0.308$ and a lookback time $T_L = 3.5$ Gyr. Here, we continue this series with comparably deep photometry in a more distant target, MACS0417.5-1154 at $z=0.443$.  Studies of this type at redshifts $z < 1$ are entirely complementary to the recent discoveries with \jwst of young GCs or proto-GCs seen at much higher redshift in the process of formation \citep{mowla+2022,mowla+2024,fujimoto+2024,adamo+2024,vanzella+2022,vanzella+2023,whitaker+2025}.  Eventually, both types of photometry will be needed to construct a complete end-to-end observational picture of GC evolution.

With standard cosmological parameters (see below), the redshift of MACS0417 converts to a lookback time of $T_L = 4.7$ Gyr, allowing us to see the physical state of its galaxies and their GC populations fully one-third of the way back to their origin. As a lensing cluster, the total mass of MACS0417 is high, near $M_{\rm vir} \simeq M_{200} \simeq 2 \times 10^{15} M_{\odot}$ as gauged through strong-lensing analysis, dynamics, the Sunyaev-Zeldovich effect, and the intracluster gas \citep{applegate+2014,bellstedt+2016,planck+2016,mahler+2019,pandge+2019}. Adopting the virial radius as
\begin{equation}
R_{\rm vir} \simeq R_{200} \equiv \Big(\frac{G M_{200}}{100 H_0^2}\Big)^{1/3} = 2.06 \textrm{Mpc} \Big(\frac{M_{200}}{10^{15} M_{\odot}}\Big)^{1/3}  
\end{equation}
then yields $R_{\rm vir} \simeq 2.6 $ Mpc for this cluster.

\begin{sidewaysfigure*}
    \centering{
    \includegraphics[width=0.99\hsize]{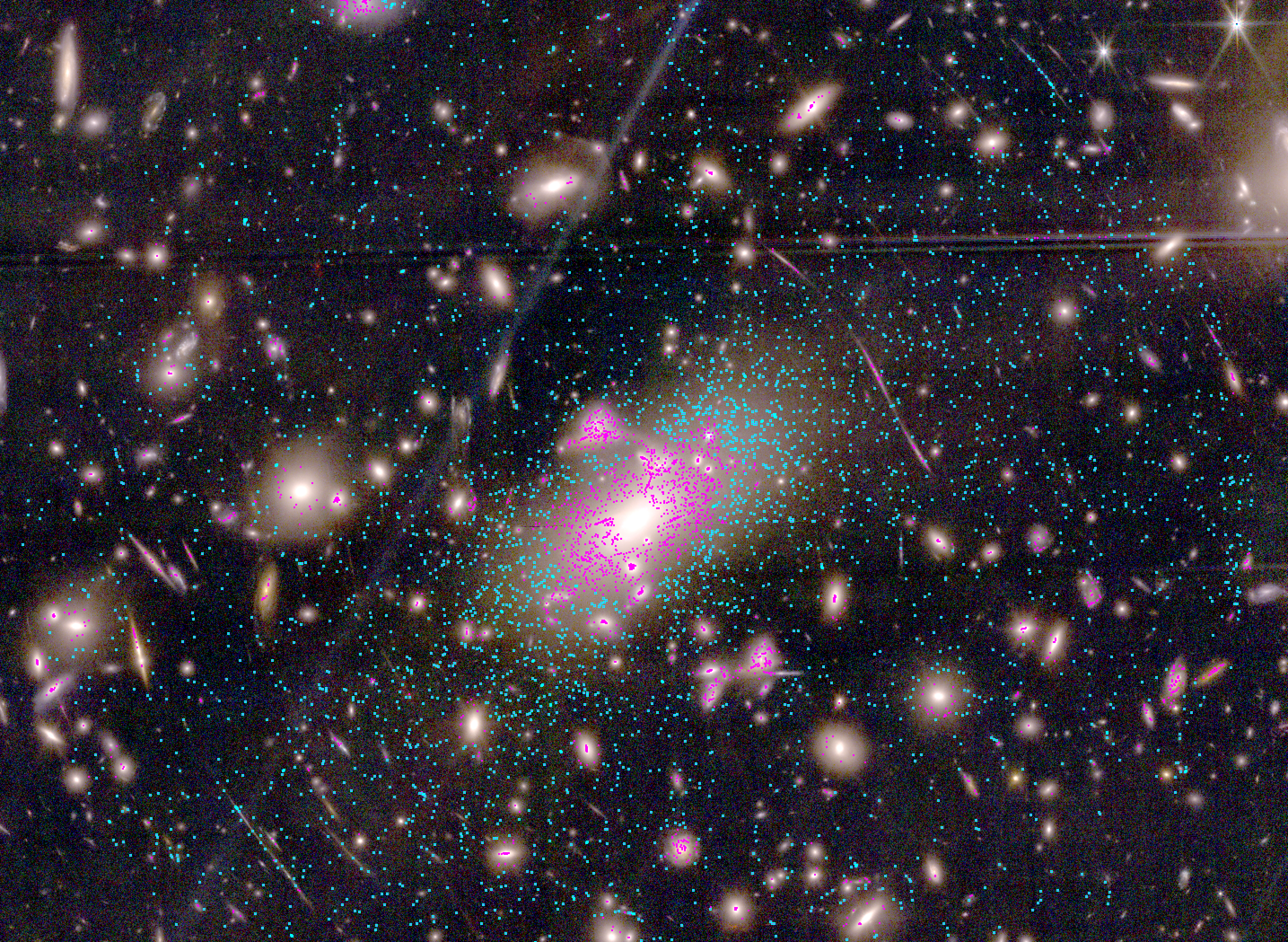}
  \caption{The central region of MACS0417, with a color image assembled from the exposures in F090W, F150W, and F200W. In this orientation, North is directed 20 degrees clockwise from vertical, East is left 20 degrees up from horizontal, and the field size shown is approximately $2.2' \simeq$ 800 kpc across. Small cyan dots show the point sources with F150W$<29.5$ in Zone 1 (see text), while magenta dots show the sources in Zone 2 defined by regions of higher local sky noise. Note the highly elongated spatial distribution around the BCG at center. The exclusion circle of $50''$ described in the text is visible from the distribution of the outer cyan points. }}\label{fig:BCGfield}
\end{sidewaysfigure*}

Structurally, MACS0417 is dominated by a BCG (Brightest Cluster Galaxy) whose stellar light profile has an unusually high ellipticity, and an orientation that closely matches the similarly elongated dark-matter potential of the cluster. The center also matches the peak of the X-ray gas emission \citep{mantz+2015,mahler+2019,jauzac+2019,okabe+2020}. In Figure \ref{fig:BCGfield}, the region of the NIRCam field centered on the BCG is shown, along with the locations of the brighter point sources (GC candidates) that were found in our study (see next section). 
In the central part of the field, traces of substructure are present within the BCG along with some active star formation, pointing to its status as a post-merger remnant \citep{pandge+2019,mahler+2019}.  Some of this substructure may be lensing-distorted background galaxies; for more detailed images see \citet{estrada-carpenter+2024}, who discuss the background ``Question Mark Pair'' of galaxies at $z=0.87$ seen just above and to the left of the BCG center, along with its five lensed images.\footnote{See also https://science.nasa.gov/missions/webb/nasas-webb-reveals-distorted-galaxy-forming-cosmic-question-mark/}

Photometry of the GC populations in galaxies in this range of redshifts is unquestionably challenging, and requires deep exposures even for \jwst. But as mentioned in \citetalias{harris_reinacampos2023}, the effect of sheer distance is partially mitigated by four other factors that work in our favor. Two of these are due to intrinsic properties of the GCs, while two others are byproducts of the redshift itself:
\begin{enumerate}
    \item Star clusters fade in luminosity with age simply because of stellar evolution, so they were more luminous in the past. The same age-fading will affect the field-star population that the GCs are superimposed on, but many GCs lie in the galactic halo where the surrounding light is low to begin with, and the additional factor of the Tolman effect on surface brightness (see item 4 below) reduces the background light further. 
\item{} All star clusters also lose mass with time due to tidal stripping and evaporation as they orbit in the potential of their host galaxies. They were therefore more massive and thus more luminous in the past. The amount of dynamical mass loss depends on both the initial GC mass and its average location relative to galactic center.
\item{} The cosmological $K$-correction, defined by $K_{\lambda} \equiv (m - M)_{\lambda} - 5~\textrm{log}(d_L/10~{\rm pc})$ where $d_L$ is the luminosity distance, becomes negative in the near-infrared spectral region for GCs. That is, the target GCs in distant galaxies are brighter in the NIR than expected only from their distance $d_L$. $K$-corrections modelled specifically for the simple stellar populations that characterize star clusters are given by \cite{reina-campos_harris2024}; for the NIRCam filters.
\item{} GCs are seen superimposed on the background light of their host galaxies, which eventually limits how far in to the galaxy center the GCs can be measured against the rising surface brightness of the central bulge. But because of the interplay of the luminosity distance $d_L = d_c(1+z)$ and the angular-size distance $d_A = d_c/(1+z)$ (where $d_c$ is the comoving distance), the surface brightness of a galaxy has a strong dependence on redshift: $SB \sim d_A^2/d_L^2 \sim (1+z)^{-4}$, independent of wavelength  \citep{hubble_tolman1935,lubin_sandage2001,condon_matthews2018}. The GCs experience the same luminosity dimming factor of $d_L^{-2} \sim (1+z)^{-2}$ as do the galaxies, but as point sources they do not have the additional $(1+z)^{-2}$ factor from the angular-size distance \citep[see also][]{windhorst+2023}. Thus at higher redshift, GCs will be projected on fainter background light, with a relative gain of $(1+z)^2$. In short, they can be detected and measured further in to the centers of the target galaxies. The same argument applies to any diffuse intracluster light within a galaxy cluster.
\end{enumerate}

The first three effects are additive, and can provide a net gain of several tenths of a magnitude at $z \sim 0.5$ relative to $z=0$.  Higher-redshift targets are certainly more challenging to reach a given absolute magnitude limit than lower-redshift ones, but not as challenging as the simple ratio of distances would at first indicate.

The outline of this study is as follows: in Section 2, the photometric measurements and artificial-star tests to assess recovery fraction are described. In Section 3, we present the color-magnitude diagrams and the luminosity function constructed from the photometry and compare the intrinsic color distributions with those of Abell 2744. In Section 4, the spatial distribution of the GCs relative to the central BCG is discussed, along with estimates of its structural parameters and comparison with the dark matter distribution. In Section 5 the CMD (color-magnitude diagram) for the MACS0417 GC system is put into direct comparison with the systems in four other BCGs at different lookback times extending up to 7.4 Gyr.  Finally, we also discuss the relative utility of the various NIRCam broaband filters for photometry of GC systems as a function of redshift. Section~\ref{sec:summary} summarizes the findings of this study.

As in \citetalias{harris_reinacampos2023}, the Planck 2015 cosmological parameters $H_0 = 67.8$ km s$^{-1}$ and $\Omega_{\Lambda} = 0.692$ \citep{planck2016} are adopted. For the MACS0417 redshift of $z = 0.443$ the luminosity distance is $d_L = 2476$ Mpc or $(m-M)_0 = 41.97$, the angular size distance is $d_A = 1189$ Mpc, and the lookback time is $\tau_L = 4.67$ Gyr.  The foreground absorption fron NED is $A_V = 0.10$ mag, equivalent to $A_{\lambda}\simeq 0.01-0.02$ in the NIRCam filters.

In the photometric measurements and analysis described below, all magnitudes are on the AB magnitude system.


\begin{figure*}
\centering{
  \includegraphics[width=0.85\hsize]{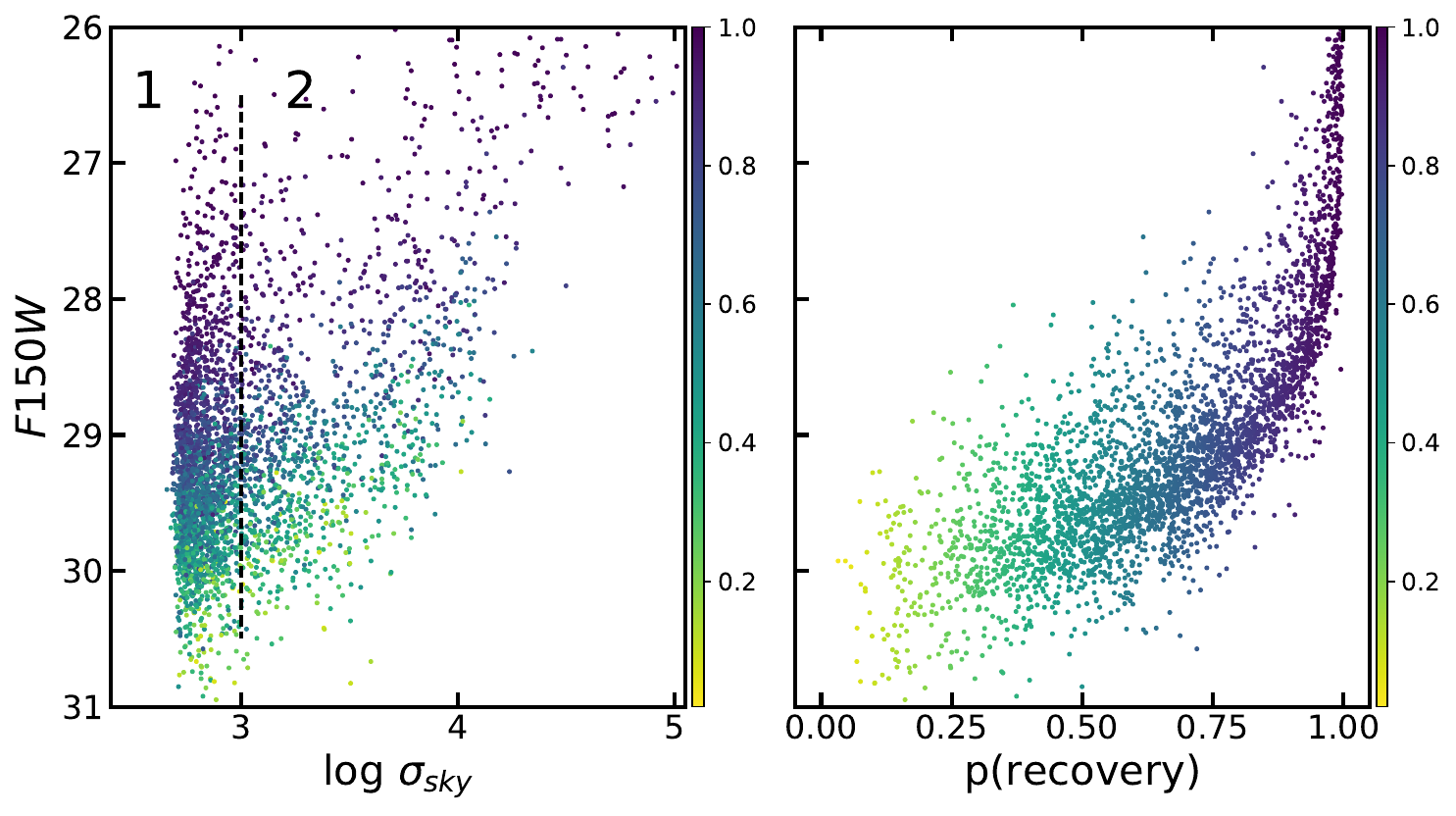}
  \caption{\emph{Left panel:} F150W magnitude versus local sky noise level for the measured GC candidates. Points are color-coded by recovery probability $p(m,\sigma_s$), with the range of $p$ shown by the colorbar at right. At higher $\sigma_s$, the effective limiting magnitude of the data becomes progressively brighter. The vertical line shows the adopted boundary for dividing the sample into Zones 1 and 2 as discussed in the text. To the left side of the line (Zone 1) the 50\% recovery probability is at F150W$_{50} \simeq 29.6$. \emph{Right panel:} F150W measured magnitude versus recovery probability $p(m,\sigma_s)$.}\label{fig:recovery}
  }
\end{figure*}

\section{Photometry and Completeness Tests} \label{sec:lr}

The data discussed in this study consist of images of the MACS0417 field taken as part of the CANUCS program (CAnadian NIRISS Unbiased Cluster Survey, JWST program 1208) of \citet{willott+2022}.  The original exposures were taken with the JWST Near Infrared Camera (NIRCam), which is capable of observing over the range 600 nm to 5000 nm with two separate cameras, the SWC (Short Wavelength Channel) and LWC (Long Wavelength Channel) that can be used simultaneously.  The maximum field size coverage for single exposures is 9.7 arcmin$^2$. For the present study, mosaic images of all exposures in the SWC filters F090W, F115W, F150W, and F200W were constructed, coordinate-registered, and resampled to a scale of 20 mas per pixel as part of the CANUCS pipeline \citep[cf.][for more details]{asada+2023,asada+2024,sarrouh+2025}. The resulting mosaics have dimensions ($20480 \times 24576$) px, corresponding to field size $410'' \times 492''$. At the distance of MACS0417 and the other lensing clusters, GCs with their typical half-light diameters of $\sim4-8$ pc are completely unresolved and thus appear as point sources on the images \citepalias{harris_reinacampos2023}. More generally, any objects with diameters less than $\sim 100$ pc will be unresolved, a range that also includes UCDs (Ultra-Compact Dwarfs). No other additional preprocessing of the images was done, such as removal of intracluster light or convolution of the images to the same effective resolution \citep{asada+2024}, in order to avoid any subtle nonlinear effects on the fluxes of the detected objects or their local background sky levels.

The first stage of measurement was to add the images of all four filters. Though this combined very deep image does not have a well defined wavelength, it is used only to obtain a finding list of candidate point sources that could then be measured on each individual filter. After this, the photometry followed the same steps used for the Abell 2744 field \citepalias{harris_reinacampos2023,harris_reinacampos2024}. The tools in \texttt{Daophot} \citep{stetson1987} within \emph{iraf} were used to run \emph{daofind} on the deep combined image with a detection threshold set to $5 \sigma_s$ where $\sigma_s$ is the standard deviation of the local sky noise. With this finding list, aperture photometry was done with \emph{phot} using an aperture radius of 3 px and a sky annulus with radii (10, 15) px. Empirical PSFs (point spread functions) were constructed from $10-15$ moderately bright, unsaturated, isolated stars on each of the four filters. The resulting PSFs have FWHM sizes of 2.1 px (F090W), 2.5 px (F115W), 3.1 px (F150W), and 3.5 px (F200W), all at the image scale of 20 mas/px. The final PSF-fitting photometry was then done with \emph{allstar}, yielding 170,000 measured objects across the entire mosaic before any culling selection was done. 

The \texttt{daophot} \emph{sharp} parameter is particularly useful for culling non-stellar objects from the total sample \citepalias{harris_reinacampos2023}. It is extremely effective in particular at removing faint, small background galaxies, as well as any objects distinctly sharper than the PSFs such as noise spikes, bad pixels, clumps not matching the PSF, or other artifacts  \citep[e.g.][]{harris+2016,harris2023}. Here the two deepest images are F150W and F200W, and these were used to isolate the sequence of star-like, unresolved sources and to reject contaminants in the candidate list. This step left approximately 59,000 objects across the mosaic. Careful inspection of the F150W and F200W images (the deepest of the four) confirmed that virtually all the point sources in the field were kept within that list, by marking all of them with small circles on a display of the field and then checking whether any point sources not falling within the circles were present.  A final culling step was done to remove objects at the centers of \textbf{about 40 of the} larger satellite galaxies, any objects near very bright stars in the field, and (the predominant source of residual contamination) any objects in the noisier regions along the detector edges. In addition, at this stage only the objects within a radius of 3200 px $=64''$ of the BCG center were kept to focus the present discussion on the central region of the galaxy cluster.

The final culled sample contains 8047 point sources that appear on the F200W and F150W images, though not all of them appear on the slightly shallower F090W and F115W images. Some small fraction of these will consist of contaminants such as nuclei of small satellites, knots along lensing arcs or spiral galaxies, foreground stars, or extremely small, faint background galaxies that lie below the resolution limit of the telescope.  Many of these in turn are removed by \textbf{the use of} local sky brightness (desribed below). Ideally, statistical removal of residual contamination would be helped by the use of an adjacent control field, but in Abell 2744 (Paper II) where a control field was available, the residual contamination was found to be at the 2\% level or less.  The result for the final CMDs in the MACS0417 field (see below) shows that the vast majority of the sample we are studying consists of the GC/UCD population.  

Special mention should be made of UCDs, which in a technical sense are also `contaminants' of the GC population but will be automatically picked up as part of the photometry and will pass through all the culling steps.  In fact UCDs are an implicit part of the study in the broader sense that we are searching for a population of point sources that belong to MACS0417. UCDs are the stellar-populations cousins of GCs, and are also routinely found in rich galaxy clusters \citep[cf.][for recent discussions]{norris+2011,chiboucas+2011,penny+2012,janssens+2017,janssens+2019,harris_reinacampos2023,pomeroy+2025}. In principle, many can be distinguished from GCs by their larger scale sizes ($r_h \simeq 10-100$ pc) and higher mass-to-light ratios \citep{hilker+2007,hau+2009,forbes+2014,janz+2015,liu+2020}. However, as mentioned above, the majority of UCDs in these distant lensing clusters will appear as point sources just as GCs do: a diameter of 100 pc would correspond to $0.017''$ angular diameter at the distance of MACS0417. Thus with only the broadband magnitudes and colors in hand, none of the UCDs can definitively be isolated as a separate subset of objects. 

Artificial-star tests (ASTs) were carried out to assess the effective photometric limits and the measurement uncertainties with the \emph{daophot/addstar} function. Approximately $6000$ fake stars covering the apparent magnitude range $27 - 32$ (the same in all filters) that generously covers the range over which the recovery fraction decreases from 100\% to zero  were inserted into the images. These were designed to follow a number density $n \sim R^{-1.2}$ relative to the BCG center, to roughly mimic the actual GC spatial distribution. The frames containing the fake stars were re-measured with exactly the same procedure and the same culling steps as for the real objects. 

The analysis of the ASTs was carried out in the same way as described in \citetalias{harris_reinacampos2024}. If a star is successfully detected and passes the culling tests, it is defined as \emph{recovered}, and the probability of recovery $p$ is modelled as a function of both magnitude $m$ and local sky noise $\sigma_s$, which directly affects the detection threshold. 
Specifically, $p(m,\sigma_s)$ is matched with a logistic regression (LR) function, 
\begin{equation}
    p = \frac{1}{1+e^{-\mathbf{g}}} = \frac{1}{1+e^{-\mathbf{X}\ \mathbf{\beta}}}.
    \label{eq:logit}
\end{equation}
where $\mathbf{g}$ is the logit function, $\mathbf{X}$ is the vector of predictor variables (in this case $m, \sigma_s$), and the $\beta_i$ are the coefficients to be solved for \citep[see also][]{rosolowsky+2021,eadie+2022,harris_speagle2024}. In principle, object crowding could be added as a third variable to the logit function \citep[see][]{harris_speagle2024}, but as noted in \citetalias{harris_reinacampos2024}, crowding and local sky noise are closely correlated, both of them increasing toward the center of the BCG or the other galaxies in the field.

\begin{figure}
\centering{
  \includegraphics[width=0.98\hsize]{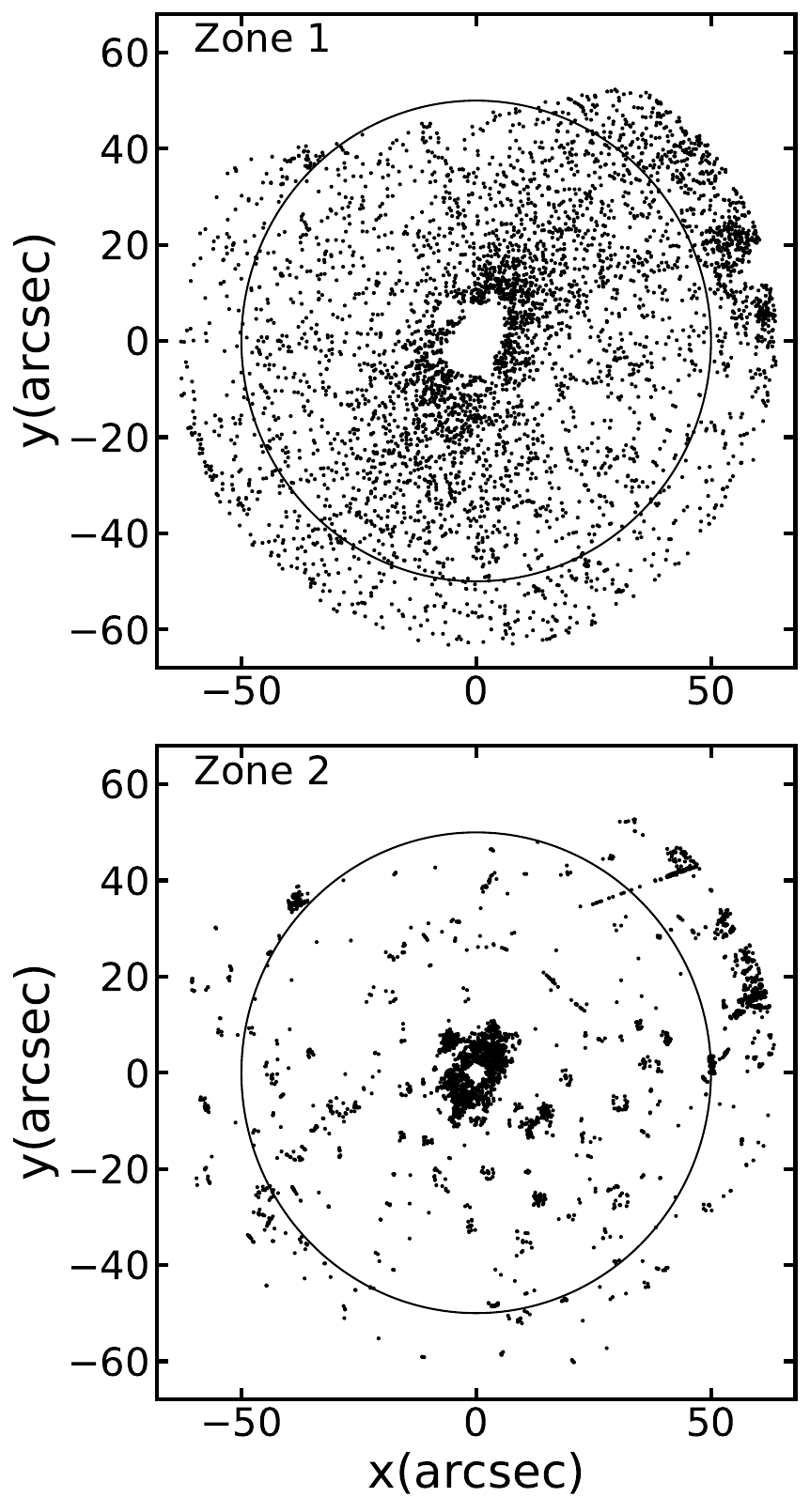}
  \caption{\emph{Upper panel:} Spatial distribution of the points sources in Zone 1 (low sky level, $\log_{10}\sigma_{s} \leq 3$). The $$(x,y)$$ values are in arcseconds and they are centered on the BCG. The circle has a radius of $50'' = 0.1 R_{\rm vir}$. \emph{Lower panel:} Point sources in Zone 2 (high sky level,  $\log_{10}\sigma_{s} > 3$).}\label{fig:xy2}
  }
\end{figure}

By contrast with the traditional approach of binning the data by magnitude and finding the average completeness of detection per bin, modelling the problem as a question of recovery probability allows us to assign a $p-$value to \emph{every individual object in the list of measurements} \citep[see][for more extensive discussion]{harris_speagle2024}.

In short, the specific logit function used here is 
\begin{equation}
    g = \beta_0 + \beta_1 F200W + \beta_2 {\rm log}_{10} \sigma_s \, .
    \label{eq:lr}
\end{equation}
where in practice $\sigma_s$ equals the standard deviation of the sky pixel intensities on the F200W image within an annulus of 7 to 15 pixels around each object, as measured with the \emph{daophot/phot} aperture photometry function. 
For the artificial stars, which cover the full desired range of magnitudes and location on the field, a maximum-likelihood solution was carried out in Python with \emph{statsmodels/scikitlearn}. The solution gives $\beta_1 = -1.80 \pm 0.04$, $\beta_2 = -1.78 \pm 0.13$. Both terms are highly significant. Further trials using the color index as a third variable \citep[see][]{harris_speagle2024} did not yield any improvement in the results.

The general effects of both variables are illustrated in two different ways in Figure \ref{fig:recovery}. The left panel shows the effect of the higher $\sigma_s$ encountered closer to the centers of the galaxies in the field, which brightens the limiting magnitude (defined as the 50\% recovery level). In the following discussion, the regions of lower local sky noise to the left of the dashed line shown in the figure (Zone 1) will be used as a `best' sample of the photometry that has the deepest limiting magnitude. In the right panel, where the magnitude in F150W is plotted directly against $p$, it is apparent that at brighter magnitudes virtually all objects are recovered regardless of location. But at fainter magnitudes, the importance of local sky noise increases dramatically and the scatter (i.e.~the range of $p-$values) becomes large.

The final set of GC/UCD candidates is also marked in Fig.~\ref{fig:BCGfield}, with Zones 1 and 2 shown in different colors.  Their spatial distribution is highly elongated, following the halo light of the central BCG, a result that will be quantified below.

The local sky level and thus the recovery probability are correlated with spatial location, as illustrated in Figure \ref{fig:xy2}. The objects in Zone 2 (high local sky noise, $\log_{10}\sigma_{s} > 3$) strongly concentrate toward the inner regions of the BCG and its smaller satellites. It is important to note that even in the inner-halo regions of Zone 2, where the candidate GC/UCDs have the highest projected number densities, they are still \emph{uncrowded}.  A good illustration of that can be seen, for example, in Figure 2 of Paper I for Abell 2744, and the same is true here.

Notably, all the point sources that survived the culling steps with \emph{sharp} but that are brighter than F150W $\simeq 26$ are nuclei of small, compact galaxies in the field, which can also be seen in Fig.~\ref{fig:BCGfield}. To concentrate the discussion further to the BCG itself, objects near two other large cluster galaxies at upper right and near the edges of the detector were avoided as well. We include for further analysis only the sample within a radius of $50'' = 290$ kpc $= 0.1 R_{\rm vir}$ around the BCG, shown by the circles in Fig.~\ref{fig:xy2}. This boundary is equal to the fiducial radius of $0.1 R_{vir}$ adopted by \citet{dornan_harris2023} as a consistent way of defining the total populations of GC systems in large galaxies.

The increase in the internal measurement uncertainty towards fainter magnitudes as determined from \emph{allstar} is accurately described by a simple exponential curve, $e(m) = e_0 + 0.1 \cdot {\rm exp}(m-m_0)$ at magnitude $m$.  The pairs of parameters are $(e_0, m_0) = (0.05, 29.3)$ (F090W, F115W), (0.04, 29.5) (F150W), and (0.04, 30.0) (F200W).

\section{CMDs and Luminosity Function}

\begin{figure}
\centering{
  \includegraphics[width=0.98\hsize]{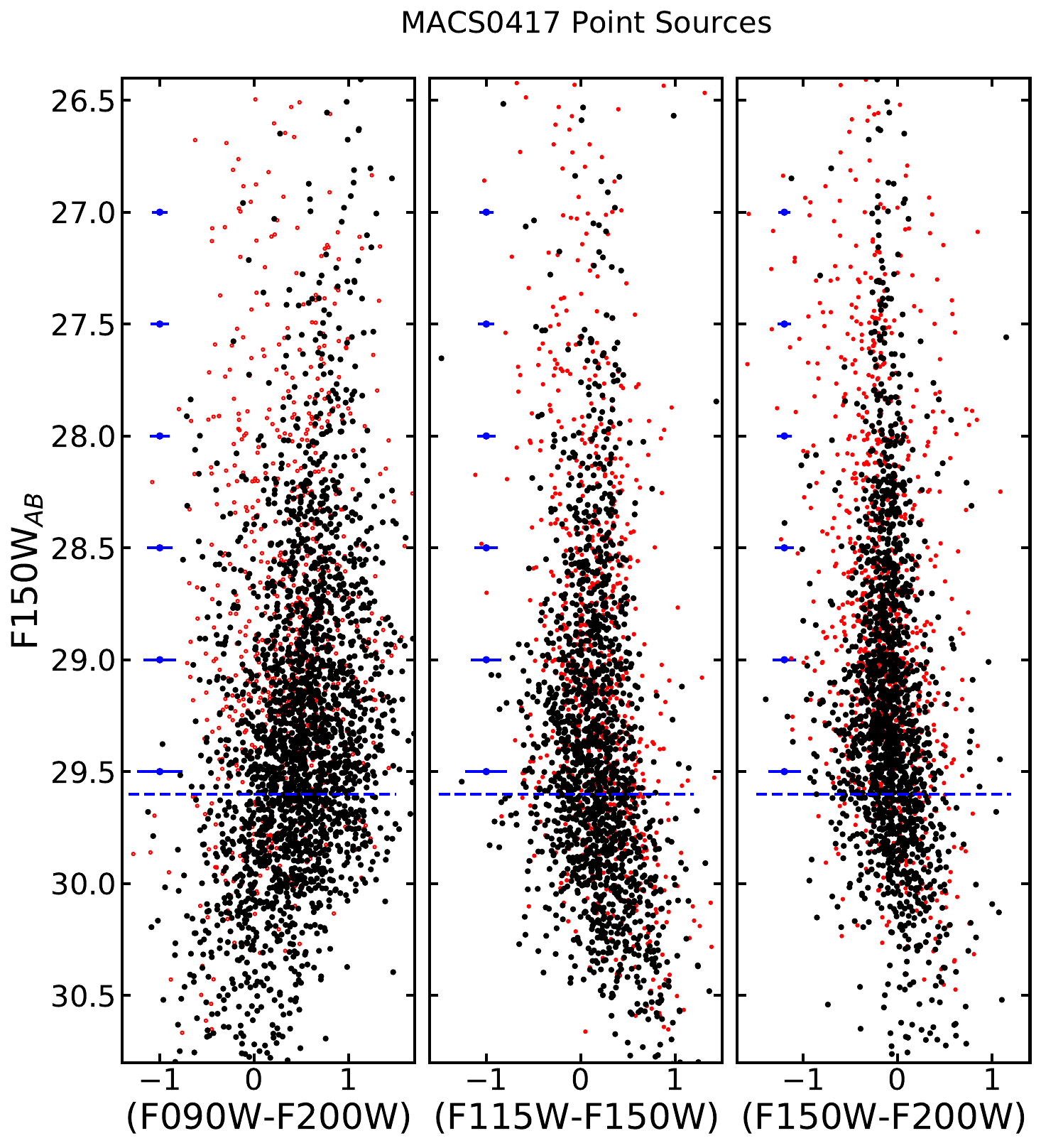}
  \caption{Color-magnitude diagrams for the culled sample as described in the text. This sample includes point sources within $50''$ of the BCG center to avoid including any parts of the other large galaxies near the edge of the image (at upper right in Fig.~\ref{fig:xy2}). Measurement uncertainties are indicated by the errorbars on the left side of each panel.  Objects in Zone 1 are plotted as black symbols, and ones in Zone 2 as red symbols. No adjustments for foreground extinction or $K$-corrections have been applied for this plot. The dashed lines at F150W $\simeq 29.6$ show the approximate 50\% recovery probability in Zone 1 (left panel of Fig.~\ref{fig:recovery}).}\label{fig:cmd3panel}}
\end{figure}

For the culled sample of point sources within $r=50''$ described above, the CMDs (color-magnitude diagrams) for three of the different color indices are plotted versus F150W magnitude in Figure \ref{fig:cmd3panel}. The (F150W-F200W) color index (third panel) is least sensitive to GC metallicity \citep{harris_reinacampos2023} and defines the narrowest sequence. In this index, the increasing spread of color towards fainter magnitudes thus best represents the photometric measurement uncertainties.

\subsection{Luminosity Function}

\begin{figure}
\centering{
  \includegraphics[width=0.94\hsize]{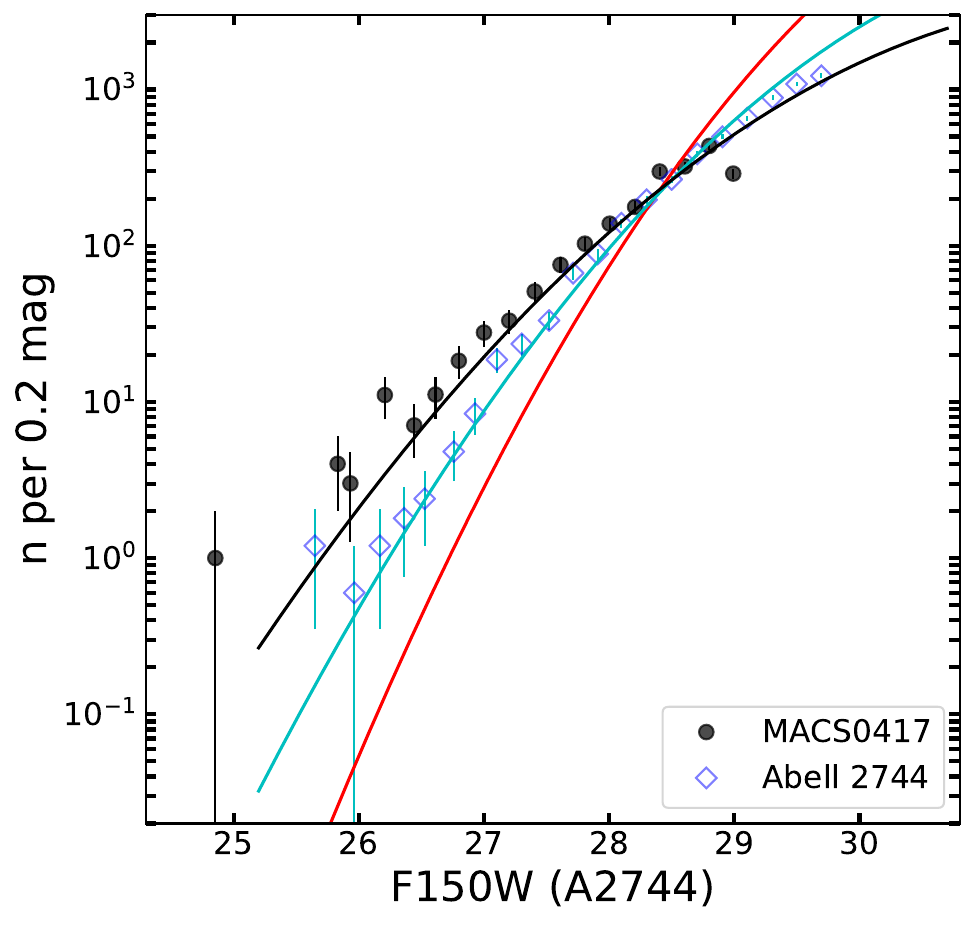}
  \caption{Luminosity function of the GC/UCD candidates, plotted as number per 0.2-mag bin in F150W, for MAS0417 (black circles) and Abell 2744 (cyan diamonds). Here the totals per bin are completeness-corrected, and the magnitudes are $K$-corrected. The MACS0417 bin magnitudes have been shifted by $-0.9$ mag to put them at the same equivalent distance modulus as Abell 2744, and the numbers are normalized to the same totals at F150W = 29.1 (see text). The solid curves show three lognormal luminosity function models with dispersions $\sigma = 1.2$ (red), 1.4 (cyan), and 1.6 (black).  }\label{fig:lf2}
  }
\end{figure}

The luminosity function of the GC candidates is shown in Figure \ref{fig:lf2}. The LF is shown in differential form as number per 0.2-mag bin in F150W. The magnitudes have been $K$-corrected with the values listed in Table \ref{tab:5systems}. The numbers shown are all completeness-corrected, and are restricted to `Zone 1' where the data reach the deepest (Fig.~\ref{fig:recovery}). The level F150W=30.0 is chosen as the faint limit to keep the mean recovery fraction per bin above $\langle p \rangle > 0.7$. Only  objects in the color range $-0.5 <$ (F150W-F200W) $< +0.7$ are used, in order to eliminate a few other outliers.

\begin{figure}
\centering{
  \includegraphics[width=0.94\hsize]{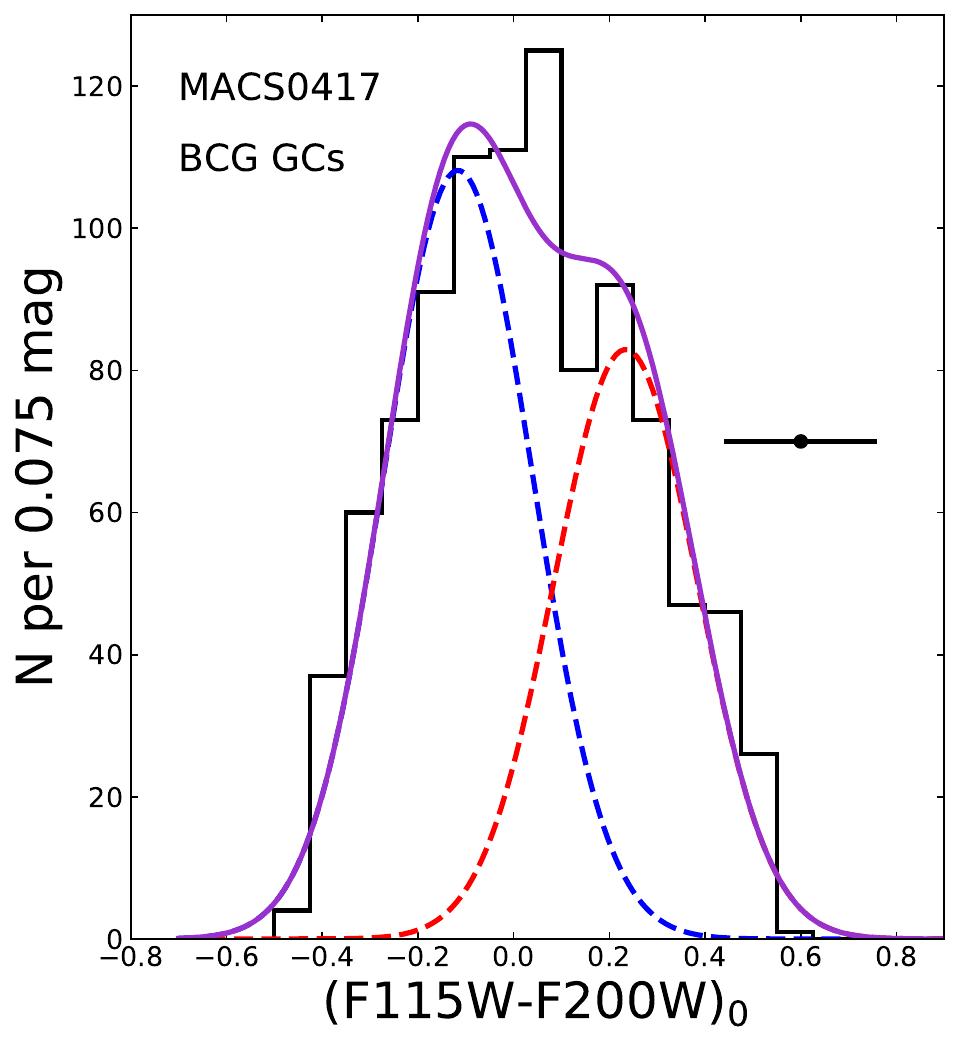}
  \caption{Distribution of color indices in (F115W-F200W), for point sources in Zone 1 and in the magnitude range $28.0 < F150W < 29.5$. The colors are dereddened and a mean K-correction has been applied (see text). A double-Gaussian fit obtained by GMM (see text) to the histogram is shown. with blue and red components plotted in the dashed lines and their sum in the solid line.  Errorbar at middle right shows the expected $1\sigma$ spread due to measurement uncertainty at F150W = 29. }\label{fig:cdf}
  }
\end{figure}

In our analysis of the Abell 2744 GC system \citep{harris_reinacampos2024}, the LF was shown to match well with a standard lognormal form in number of objects per unit magnitude,
\begin{equation}
\phi(F150W) = \frac{N_{\rm tot}}{\sqrt{2 \pi} \sigma} \exp\left[\dfrac{-(F150W-m_0)^2}{2 \sigma^2}\right] \, .
\label{eq:lf}
\end{equation}
For GCs fainter than the peak or turnover point at $F150W = m_0$, the EMP-Pathfinder models \citep{reina-campos+2022b} would predict that the LF will become broader at earlier stages of evolution where the dynamical effects on lower-mass clusters are not as advanced; but these faint levels are not reached in the present observational data. Thus the LF dispersion $\sigma$ in Eq.~\ref{eq:lf} above refers only to the bright half of the distribution.

The photometry in MACS0417 does not reach the same GC luminosity limit as for Abell 2744, both because the total exposure times of the original images are shorter, and because MACS0417 is more distant by 0.9 magnitude. Thus here, no attempt is made to fit the LF by  solving simultaneously for the turnover $m_0$ and dispersion $\sigma$, since for data that fall well short of the turnover, these two parameters are highly correlated \citep{hanes_whittaker1987,harris+2014}. Instead, we simply assume that the  LF turnover point is at $m_0 = 33.1$, i.e. 0.9 mag fainter than the value $m_0 =32.17$ found for Abell 2744 \citep{harris_reinacampos2024}. Then the dispersion is estimated by matching the differential LF to different values of $\sigma$, requiring $m_0$ to be the same in all cases. As Fig.~\ref{fig:lf2} shows, $\sigma \simeq 1.6$ matches well, a value about 0.2 mag larger than the typical observed dispersions for BCG-type galaxies in the nearby universe \citep{villegas+2010,harris+2014}.

These parameters can also be used to make a rough estimate of the total GC population in MACS0417. There are $2007 \pm 52$ objects brighter than F150W = 30.0 (including completeness correction) in Zone 1, which falls $1.94 \sigma $ short of the assumed turnover and thus includes just $2.6\%$ of the total population \emph{assuming} that the LF continues downward to follow the lognormal shape. Extrapolating this function over all magnitudes then gives $N \sim 76,600$ GCs within the area of Zone 1. The uncertainty is perhaps $\pm 30,000$ given that both the assumed LF dispersion and turnover level are uncertain by $\pm 0.1$ mag. Going beyond this to add in the area of Zone 2, as well as the other MACS0417 galaxies not included in this survey, will increase this total by perhaps a factor of two, using the comparable material in Abell 2744 as a guide (Paper II).  Thus we estimate very roughly $N_{GC}(tot) \sim 1.5 \times 10^5$, recognizing the large extrapolation required. For comparison, in Abell 2744 the estimated total was $N(tot) \simeq 1.1 \times 10^5$ GCs. 

An additional \emph{caveat} is that these extrapolations assume the lognormal LF shape is symmetric, so these estimated totals may best be thought of as the predicted numbers that these two galaxy clusters would have at the present epoch, after the fainter half of the LF has been subjected to 3--4 more Gyr of dynamical evolution and many more low-mass GCs have been removed.

Perhaps a more effective (and model-free) comparison is to match the A2744 and MACS0417 GCLFs directly. Fig.~\ref{fig:lf2} shows this comparison, where the MACS0417 LF has been shifted by $-0.9$ mag to put it at the same distance modulus as A2744. The A2744 LF (cyan symbols) is directly superimposed on it, normalized to the same total number brighter than F150W = 29.1, and also shifted horizontally by $-0.2$ mag to adjust for the age fading due to stellar evolution over the 1.2 Gyr difference in lookback time (see next section below and Fig.~12 of Paper I).

The A2744 GCLF is clearly steeper at these high luminosities, matching a dispersion of $\sigma \simeq 1.4$ (see also Paper II) that is much more similar to the results for nearby BCGs \citep{harris+2014} instead of the $\sigma \simeq 1.6$ characterizing MACS0417. The same trend is seen in the cumulative LF of the lower panel in Fig.~\ref{fig:lf2}. 
For levels fainter than F150W $\gtrsim 28$, the raw data do not permit clear conclusions about the LF shape and potential similarity between the two clusters. Photometry reaching at least a magnitude deeper will be needed for that purpose. A more cautious way to state the current findings is that in MACS0417, seen at a less dynamically evolved state, the LF shape in the `superluminous' GC or UCD regime is shallower. Additional comments will be made in the next section, but a potentially related observation, raised by \citet{pfeffer+2025}, is that in recent JWST imaging studies of extremely young star clusters at high redshift, there are more very high-luminosity objects than are predicted in contemporary simulations \citep[see][]{pfeffer+2025}. These hints suggest that further observations of GC systems at redshifts of 0.5 and beyond may reveal intriguing trends at the high-luminosity (and presumably high-mass) end that were not suspected from the local Universe.

\subsection{Color and Metallicity Distribution}

The intrinsic metallicity distribution function (MDF) of the GCs is reflected in their color distribution function (CDF). In principle we would like to know if, for example, the MDF has the classic bimodal or multimodal shape as seen in the intrinsic color distribution for large galaxies \citep[e.g.][]{ashman_zepf1995,peng+2006,cho+2016,hartman+2023,harris2023}.  Though colors for GCs many Gyr old are insensitive to age, the essential observational difficulty in converting colors to metallicity is that NIR color indices are much less sensitive to metallicity when compared with those in the optical bands \citep{cho+2016,hartman+2023,harris2023,harris_reinacampos2023}.  For our MACS0417 data the best choice for the combination of metallicity sensitivity and photometric measurement uncertainty is the (F115W-F200W) index, which was used for Abell 2744 (Paper I).  

For the present study we will use the CDF for a preliminary estimate of the MDF, with more complete discussion based on simulated samples of GCs to be carried out in a followup study. To generate a CDF the point sources in the CMD in the magnitude range $28.0 < F150W < 29.5$ are used to exclude both objects fainter than the completeness limit at the faint end, and possible UCDs at the bright end which on average will populate the high-luminosity end preferentially.  Results are shown in Figure \ref{fig:cdf}; overall, the appearance of the CDF is very similar to the result obtained for Abell 2744 (Paper I). The distribution contains 976 points, and can be well matched with a standard double-Gaussian function.   The fit shown in the figure was done with the mixture model code GMM \citep{muratov_gnedin2010}, yielding the parameters listed in Table \ref{tab:cdf}.  Although the scatter due to simple measurement uncertainty is important (see errorbar in figure), the observed range in (F115W-F200W) is about three times larger, so the intrinsic width of the CDF is resolved.  A single-Gaussian fit is rejected at $>$99\% significance.  The raw color indices have been dereddened, and K-corrected assuming an age of 7 Gyr and metallicity [m/H] $= -0.96$. The two fitted components (blue, metal-poor versus red, metal-rich) have roughly equal numbers of objects (though with significant uncertainty), which is typical for GC systems for giant galaxies in the local universe \citep{peng+2006,hartman+2023,harris2023}.  

The metallicity sensitivity is $\Delta {\rm [m/H]} \simeq 5 \Delta C$ where $C = $(F115W-F200W) (see Paper I).  The main systematic uncertainty in converting the peak colors of each component, however, is in the K-corrections: the K-values depend little on GC age, and have only a shallow dependence on metallicities at \emph{low} metallicity.  But at metallicities higher than [m/H] $\simeq -0.5$, they start changing more steeply with metallicity. The intrinsic corrected color indices  are uncertain by $\pm0.02$ mag at the low-metallicity end from the K-corrections alone, but this increases to more than $\pm0.1$ mag at the metal-rich end.  Adjusting for this effect, and using the metallicity/color conversion from Paper I, we estimate that the mean metallicity of the blue-GC component is [m/H] $\simeq -1.0$ and the red-GC component is near [m/H] $\simeq -0.3$.  Keeping in mind the serious uncertainties built into this conversion, these values are $\sim0.2$ dex more metal-rich than those found for nearby giant galaxies \citep{peng+2006,hartman+2023,harris2023}.

A better approach, to be pursued in followup work, will be to use simulated GC systems to determine K-corrections and color indices for clusters as a function of both their age and metallicity individually, and then simulate the entire CMD. Paper II contains a first example of this method, employing the EMP-Pathfinder models  \citep{reina-campos+2022b}. EMP-Pathfinder is a suite of cosmological zoom-in simulations of galaxy formation for Milky-Way-sized galaxies, wnich specifically follow the formation and dynamical evolution of star clusters within them.  At any given epoch (redshift), the existing clusters are all tagged with their individual ages, masses, and metallicities, which can then be transformed into the observational plane of luminosity and colorm and finally matched up with the observed distribution in the CMD. 

\begin{table}
\centering{
\caption{Parameters for the Color Distribution Function} \label{tab:cdf}
\begin{tabular}{cccc}
 \hline \hline
Component & Mean & Dispersion & Fraction  \\
\hline
Blue & $-0.116 \pm 0.083$ & $0.155 \pm 0.037$ & $0.524 \pm 0.243$\\ 
Red  & $+0.234 \pm 0.104$ & $0.150 \pm 0.041$ & $0.426 \pm 0.243$ \\
\hline
 
\end{tabular}}
\end{table}

\section{Spatial Structure}

\begin{figure}
\centering{
  \includegraphics[width=0.94\hsize]{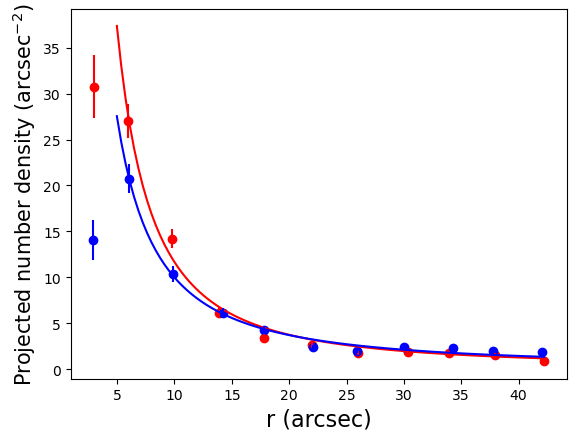}
   \includegraphics[width=0.94\hsize]{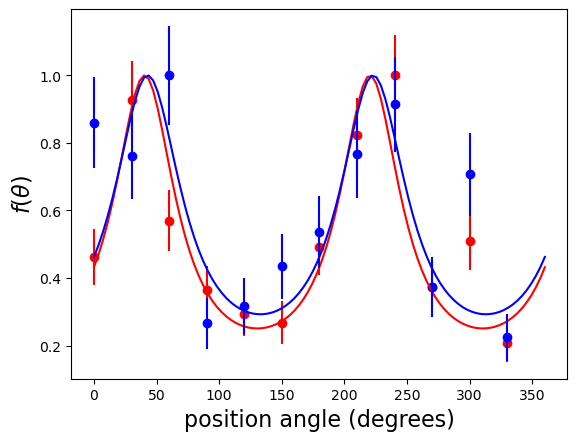}
  \caption{\emph{Top panel:} Radial distribution of point sources around the BCG. Objects in the bluer and redder halves of the CMD in (F115W-F200W) are plotted separately. \emph{Lower panel:} Azimuthal distribution in 20-degree sectors, again separately for the bluer and redder GC subsets.}\label{fig:shape}
  }
\end{figure}

In many large galaxies that hold GCs over a wide range of metallicities, the `bluer' more metal-poor clusters are seen to follow a more extended spatial distribution than the `redder' more metal-rich ones and there are very roughly equal numbers of GCs in both subsets \citep[e.g.][]{geisler+1996,harris2009,liu+2011,forbes_remus2018,harris2023}. Although the NIR color indices used here are not highly sensitive to metallicity, and the spread of colors in the CMDs (Fig.~\ref{fig:cmd3panel}) further blurs any clear separation between the blue and red populations, an approximate test can be made of this trend simply by dividing our sample in half and plotting each half separately.

In Figure \ref{fig:shape} (upper panel), the radial distribution of point sources for the high-completeness range F150W $< 29.6$ is shown as number density $\sigma$ versus radius.  Here $r$ is defined in circular annuli, so effectively $r = \sqrt{ab}$ for a simple elliptical shape. A power-law fit $\sigma \sim r^{-\alpha}$ to the datapoints (shown in the figure) gives an exponent $\alpha = 1.60 \pm 0.06$ for the entire system, $1.47 \pm 0.05$ for the blue subset, and $1.68 \pm 0.09$ for the red subset. The innermost radius shown is expected to be reduced by incompleteness more than the outer points (see above) and is not used for these fits. The points clustered around a large satellite galaxy to the upper left of the BCG were also ignored. 

The azimuthal (angular) distribution provides an additional test of the blue/red differences and also the alignment of the GC distribution with the BCG halo light. The lower panel of Fig.~\ref{fig:shape} shows the results, where the points are divided into 20-degree sectors within the same outer radial limit used for the radial fit, $r<50'' = 290~$kpc. Matching the azimuthal distribution with the method derived in \citet{mclaughlin+1994} gives an eccentricity (1) $e = b/a = 0.55 \pm 0.03$ and position angle $\theta = 42 \pm 3$ deg N of W for all points, (2)
$e = 0.54 \pm 0.05, \theta = 43 \pm 5$ deg for the blue objects, and (3) $e = 0.58 \pm 0.03, \theta = 41 \pm 3$ deg for the red point sources. For comparison, the extended halo light of the BCG has $e = 0.46$ at a radius of $20~$kpc and $\theta = 49$ deg N of W \citep{jauzac+2019,okabe+2020}, in good agreement with the GC distribution. 
In summary, we find no significant difference between the blue and red GC subgroups in either their radial or angular distributions: both of them define a highly flattened halo system with the same orientation as the halo light. 

A comparison can also be made with the more extended dark-matter (DM) potential evaluated through the gravitational lensing map, for which $e=0.67, \theta = 56$ deg \citep{okabe+2020}. The DM distribution has a similar orientation angle but defines a still more flattened elliptical shape than does the BCG itself and its GCs.

The scale size of GC systems in galaxies, gauged either by the power-law index $\alpha$ or the effective radius $R_e$, increases with galaxy luminosity \citep[e.g.][]{chen_gnedin2024}.  In lower-luminosity galaxies the GC distribution is generally more extended than the underlying galaxy light (particularly for the metal-poor GCs), but for BCGs, the GC radial distribution converges toward the underlying halo light and the DM distribution primarily because of their higher proportion of `ex situ' stars and clusters brought in by mergers and satellite accretions \citep[][and references cited above]{forbes_remus2018,choksi_gnedin2019,reina-campos+2022,chen_gnedin2024}.

\section{A Synoptic Picture of GCS Evolution}

Enough material is now available through recent photometry of GC systems in giant lensing clusters of galaxies that we can begin, for the first time, to assemble a \emph{purely observational} look at the nature of these systems at different stages of their evolution, extending over a timespan of many Gyr. 

\subsection{GC Systems Viewed Over Cosmic Time}

\begin{figure*}
\centering{
  \includegraphics[width=0.95\hsize]{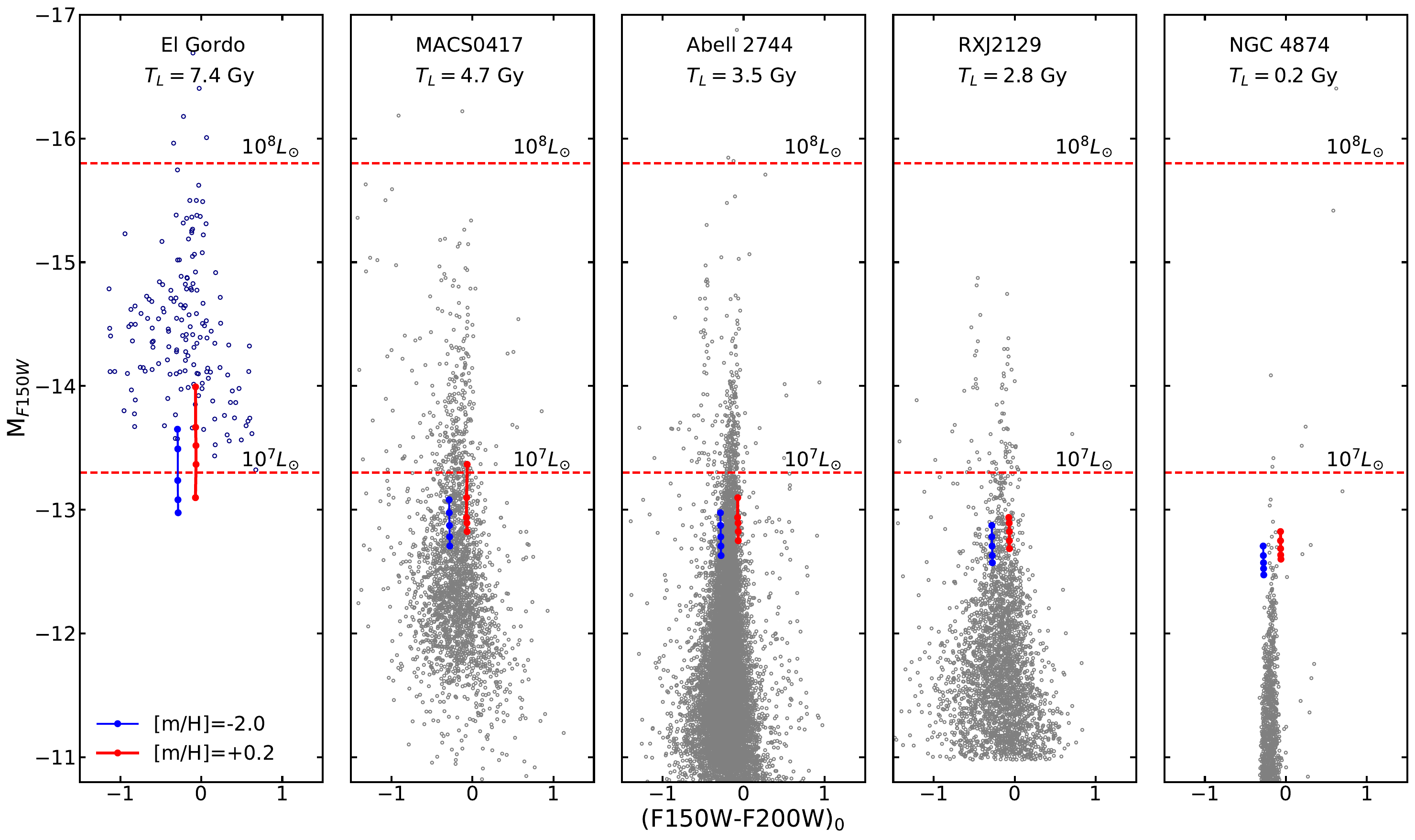}
      \caption{CMDs for the GC/UCD populations in five massive galaxy clusters over a range of redshifts, drawing from recent JWST and HST measurements. They are arranged in order of age (reverse order of lookback time). Upper and lower dashed lines mark fiducial luminosity levels $10^7 L_{\odot}$ and $10^8 L_{\odot}$. The blue and red dots and lines in each panel mark the predicted luminosities of a $2 \times 10^7 M_{\odot}$ GC at two different metallicities (--2.0 and +0.2) and different ages appropriate for the given lookback time (see Table~\ref{tab:5systems} and text). For each red/blue set of points the age increases from top to bottom. The decrease in luminosity from one panel to the next for these model lines is due only to direct stellar evolution, not accounting for any GC mass loss. Note that for the nearest object (NGC 4874 in the Coma cluster) the (F150W, F200W) magnitudes have been transformed from the HST optical bands (F475W, F814W); see text.  \emph{NB:} In A2744 and RXJ2129, the thin sequences to the blue side of the GC/UCD sequence at a color index (F150W-F200W)$_0 \simeq -0.5$ appear to be objects unassociated with the BCGs; see Paper I for discussion. }\label{fig:cmd5}}
\end{figure*}

As a first of its kind, Figure \ref{fig:cmd5} shows the CMDs of five target galaxy clusters at different cosmic epochs, placed in direct comparison with each other. The sources of their photometry are listed in Table \ref{tab:5systems}. Here, intrinsic luminosity $M_{F150W}$ is plotted versus color (F150W-F200W)$_0$, now including the $K$-corrections along both axes. Five distinct lookback times $T_L$ are represented. For the present work, we restrict our comments to a few general interpretations, deferring more quantitative comparisons with theory to a later discussion.

\begin{table*}
\centering{
\caption{Parameters for Five Massive Clusters} \label{tab:5systems}
\begin{tabular}{ccccccl}
 \hline \hline
Target System & $(m-M)_0$ & $z$ & $T_L$ (Gyr) & $K_{F150W},K_{F200W}$ & GC Age Range (Gyr) & Source \\
\hline
NGC 4874 & 35.25 & 0.02 & 0.2 & $-0.02,-0.04$ & $9 - 13$ & \citet{harris2023} \\ 
RXJ2129 & 40.43 & 0.23 & 2.8 & $-0.13,-0.37$ & $7 - 11$ & \citet{keatley_harris2025} \\
Abell 2744 & 41.06 & 0.30 & 3.5 & $-0.17,-0.42$ & $6 - 10$ & \citet{harris_reinacampos2023}\\
MACS0417 & 41.97 & 0.44 & 4.7 & $-0.20,-0.47$ & $5 - 9$ & This study \\
El Gordo & 43.79 & 0.87 & 7.4 & $-0.08,-0.47$ & $2 - 6$ & \citet{harris+2025} \\
 \hline
 
\end{tabular}}
\end{table*}

NGC 4874, the BCG in the Coma cluster, requires special comment. Data for GCs in several lensing clusters at various redshifts are now in hand or in progress, all through JWST/NIRCam imaging. But somewhat paradoxically, no such material is available for any local BCGs to give us a ``zero redshift'' anchor for comparison with the redshifted systems. As a temporary step, we show here the GC photometry for the rich NGC 4874 GCS from \cite{harris2023}. The data were obtained in (F475W, F814W) with HST/ACS, but now transformed into (F150W, F200W) in the NIRCam AB system. Since no purely empirical transformations between these two sets of bands are available, we have employed the PARSEC CMD3.7 stellar models \citep{bressan+2012} for this purpose. The adopted conversions are:
\begin{align}
 &  (F150W - F200W) \simeq -0.55 + 0.345 (F475W-F814W), \\
 & F150W \simeq F814W + 0.43 - 0.86 (F475W-F814W), 
\end{align}
where the NIRCam indices are in the AB system and the HST/ACS indices are in Vegamag. The (F475W-F814W) color index is far more sensitive to GC metallicity than (F150W-F200W) is \citep[cf.][]{harris2023,harris_reinacampos2023}, so the effect of this transformation is to compress the GC color range into quite a narrow sequence.

These same PARSEC models can be used to do a more physically instructive comparison, using the fact that we are seeing five GC systems at distinctly different mean ages. We can expect that the \emph{age range} among the GCs in a given system will be at least 4 Gyr, if their formation times span the same range as they do in the Milky Way and M31, and that higher-metallicity GCs on average form later than lower-metallicity ones \citep{vandenberg+2013,leaman_vandenberg2013,forbes2020,kruijssen+2020,usher+2024,joschko+2024}. For purposes of illustration in Fig.~\ref{fig:cmd5}, the PARSEC stellar models have been used to plot the predicted (F150W-F200W) intrinsic colors and luminosities for GCs in NGC 4874 at ages of 9, 10, 11, 12, and 13 Gyr. These are shown as the blue and red connected dots in the last panel of Fig.~\ref{fig:cmd5}. For convenience, the adopted mass for all GCs is $2 \times 10^7 M_{\odot}$, near the maximum observed mass for GCs in the local Universe \citep{harris+2014,harris2023}. The colors are calculated for two metallicities, [m/H] = $-2.0$ (blue dots) and +0.2 (red dots).

The pairs of blue and red lines and dots in the other panels show the luminosities and colors \emph{for the same model GCs}, at $M = 2 \times 10^7 M_{\odot}$ and metallicities of $-2.0$ and +0.2, and over the same 4-Gyr age range, but with absolute ages reduced by the lookback time $T_L$ of each target system. Table \ref{tab:5systems} gives the specific ranges used. The steady decrease in mean luminosity from the leftmost panel to the rightmost one shows the result of simple age-fading due to stellar evolution. It does not include any additional luminosity decreases due to mass loss, although for GCs above $\sim 10^7 M_{\odot}$ the dynamical mass loss will be relatively small \citep[e.g.][]{li_gnedin2019}.

If the photometric calibrations for each target are correct; and if the $K$-corrections are correct; and if the predicted model luminosities and colors are correct, then each pair of blue/red lines should bracket the observed GC sequence. At progressively fainter magnitudes, photometric measurement scatter will broaden the observed sequences, but the fiducial lines are well centered on the sequences for all except the El Gordo data, where the observed sequence lies closer to the red-GC line. In this highest-redshift case, the $K$-corrections become more sensitive to age and metallicity and are uncertain by $\sim 0.1$ mag; detailed discussion of the El Gordo material is in \citet{harris+2025}.

The most immediately visible trend across the redshift sequence is the decrease in luminosity of the GC sequences as redshift decreases, mimicking the red/blue fiducial lines. In short, \emph{the expected age-fading of the sequence is, for the first time, directly visible through the compilation of this observational sequence}. Three other effects that are evident from the model lines are that (1) GC luminosity becomes less sensitive to age at lower redshift (the red/blue lines showing the 4-Gyr age interval are shorter); (2) the color index remains virtually constant with age (the lines are vertical); and (3) in these near-IR indices the higher-metallicity GCs reach higher luminosity than do the lower-metallicity ones (the red lines sit higher than the blue ones), though the amount of this offset decreases with age.

Notably, in these lensing clusters there are significant numbers of points lying well above the $2 \times 10^7 M_{\odot}$ level marked by the fiducial pair of lines.  Though some of this may be due to differences in the total GC populations present in the different systems, we suggest that many of these super-luminous objects may be UCDs, 

\citet{norris+2011} in particular suggest that up to a mass of $7 \times 10^7 M_{\odot}$ (roughly $M_{F150W} \sim -14.7$ in Fig.~\ref{fig:cmd5}, or $M_{F150W} \sim -15$ for the much younger El Gordo system) the population is a mix of super-luminous GCs and stripped dwarf nuclei (GCs versus true UCDs). This range would cover almost all the high-luminosity part of the sequences in the five systems of Fig.~\ref{fig:cmd5}. In this respect, it is notable as well that the LF discussed in the previous section shows no special break point at high luminosity marking any clear changeover to a different type of population \citep[see also][]{pomeroy+2025}.

\begin{table}
\centering{
\caption{Fractional Ratio of UCDs} \label{tab:ucdfraction}
\begin{tabular}{ccccc}
 \hline \hline
Target System & Threshold $M_{150}$ & $N_{hi}$ & $N_{lo}$ & Ratio \\
\hline
RXJ2129 & $-12.94$ & 119 & 494 & $0.24 \pm 0.03 $ \\
Abell 2744 & $-13.10$ & 340 & 1256 & $0.27 \pm 0.02$ \\
MACS0417 & $-13.37$ & 237 & 834 & $0.28 \pm 0.02$ \\
 \hline
\end{tabular}}
\end{table}

The NGC 4874 field is again an exception, showing little or no UCD contribution. This is a selection effect, because at its distance of just 107 Mpc, objects larger than $\sim 10$ pc radius are easily resolved and identified as non-point sources. These non-stellar objects were systematically removed by the culling process during the photometric analysis, as described above \citep{harris2023}. However, they are present in the Coma system measurements by \citet{pomeroy+2025}, and the cluster-wide Coma UCD members are discussed separately by, e.g., \citet{chiboucas+2011,pomeroy+2025}

Even ignoring NGC 4874 and El Gordo, the number of datapoints in the high-luminosity (UCD) range increases with lookback time. Once adjusted for evolutionary effects, however, the trend is largely removed. We can quantify the effect very roughly by taking the number of points in the CMDs lying above the $2\times 10^7 M_{\odot}$ level in each panel, and dividing by the number of points in the next one-magnitude interval below that level. In this way, the counts are restricted to the range unaffected by photometric incompleteness, and will also compensate for luminosity evolution. The results of this exercise are shown in Table \ref{tab:ucdfraction}. The table lists the threshold $M_{150}$ dividing line, the numbers of objects $N_{hi}$ and $N_{lo}$ above and below that line, and the ratio $N_{hi}/N_{lo}$ where the quoted uncertainty is from the Poisson count statistics. Outliers with extreme colors were removed from the counts. In all three systems, the `UCD fraction' ($N_{hi}/N_{lo}$) remains nearly uniform, increasing very slightly with redshift though not significantly. Although this result may seem at odds with the shallower LF slope for MACS0417 discussed above, the simple counts in Table \ref{tab:ucdfraction} do not take into account the details of the distribution in magnitudes.

The most important gap in the sequence shown in Fig.~\ref{fig:cmd5} is between MACS0417 and El Gordo, a step of 2.7 Gyr in lookback time during an epoch when GC systems are still rapidly evolving. New observations of target lensing clusters near $z \sim 0.7$ are much needed now to help trace out the trends at the high-luminosity end of the GC/UCD sequence that are only hinted at in the material available at present.

A more detailed modelling of the combined effects of stellar evolution and mass loss, and comparison with simulations of GC systems, is an extremely interesting follow-up step and will be deferred to a later discussion, when additional deep photometry for more lensing clusters will also be available.

\subsection{Which Camera and Filters to Choose?}

For a final part of our discussion, we raise a question central to the ongoing photometric studies of GC systems in distant targets, which is to ask which filters and what wavelength regions will be most effective. There are two competing effects at work here. First is that the SED (spectral energy distribution) for any individual GC will shift to longer wavelength at higher $z$, arguing that redder filters would be more useful. But the second, opposite effect is that these remote GCs are seen at younger ages simply because of their lookback time, which shifts the SED to the blue because their light is coming from younger stars. Which effect wins, or do they effectively cancel?

\begin{figure*}
\centering{
 \includegraphics[width=0.95\hsize]{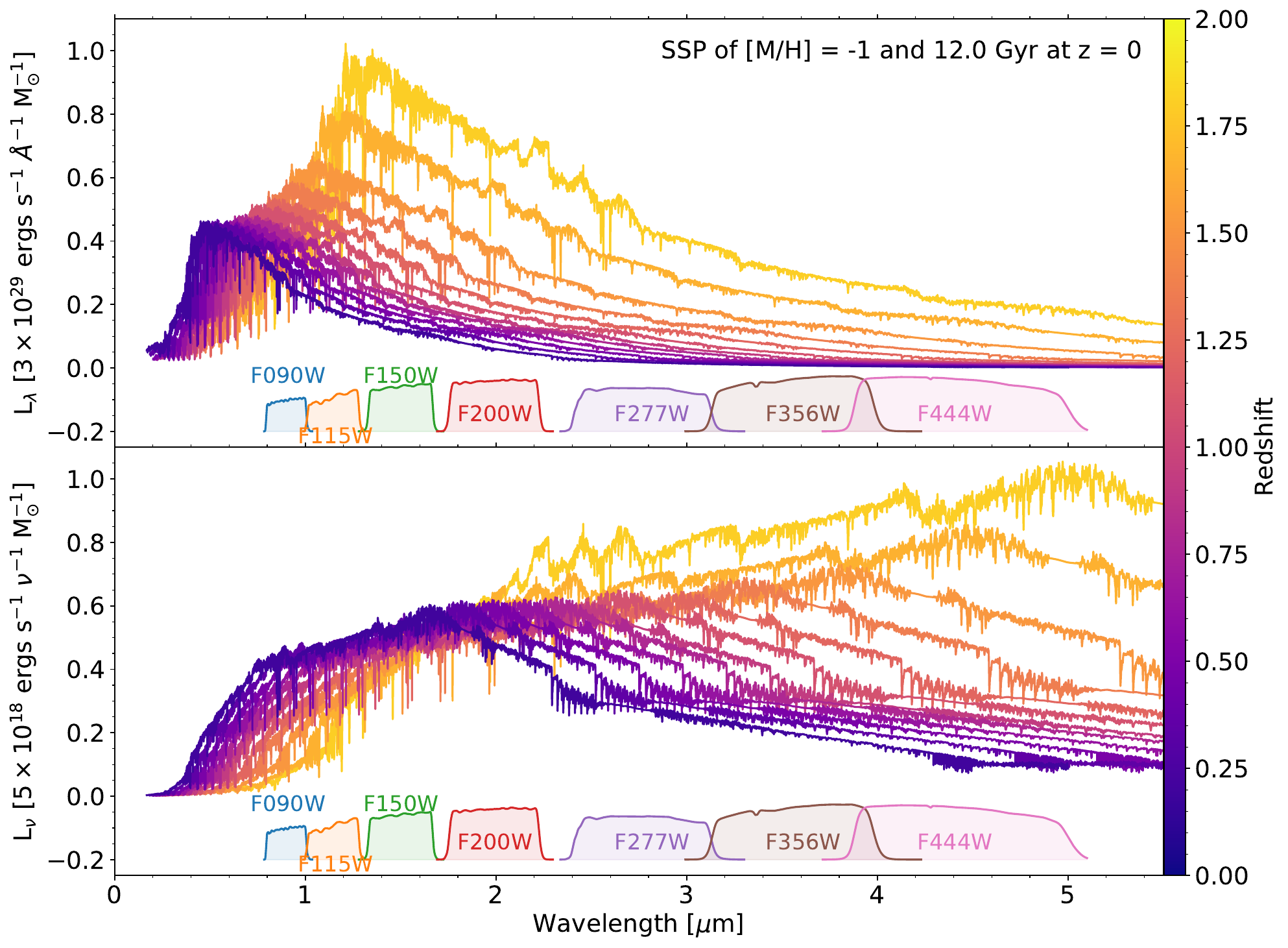}
  \caption{\emph{Upper panel:} Model SEDs showing luminosity $L_{\lambda}$ versus wavelength for a SSP from E-MILES, at increasing redshifts. \emph{Lower panel:} Model SEDs now plotted as luminosity $L_{\nu}$ versus wavelength. Successive redshifts (from left to right) are $z = 0.0, 0.1, 0.2, 0.3, 0.4, 0.5, 0.6, 0.8, 1.0, 1.3, 1.7, 2.0,$ Each model curve has an age equal to $(12 - T_L(z))$ Gyr where $T_L(z)$ is the lookback time corresponding to $z$. These curves may be used to gauge the wavelength range and thus the filters across which the detected signal from the GC is maximal.}\label{fig:seds} }
\end{figure*}

Figure \ref{fig:seds} illustrates the expected trend for a series of sample SEDs over the redshift range $0 < z < 2$. The upper graph shows SED luminosity per unit wavelength $L_{\lambda}$, while the lower graph shows the same SEDs as luminosity per unit frequency $L_{\nu} \sim \lambda^2 L_{\lambda}$. Each plotted curve shows the \emph{same GC} seen at different redshifts and thus at different ages in its history. The fiducial model SED is from the E-MILES library \citep{rock+2016} for a Simple Stellar Population of metallicity [m/H] $= -1$ and an age of 12 Gyr at zero redshift (that is, an average GC in the nearby Universe). Successive curves color-coded by $z$ are its redshifted SED for $z = 0.0, 0.1, 0.2, 0.3, 0.4, 0.5, 0.6, 0.8, 1.0, 1.3, 1.7, 2.0$. By definition, the age of each model is $(12 - T_L(z))$ Gyr where $T_L(z)$ is the lookback time.

The net result is that the redshift evolution is slightly the more important of the two competing effects. But for the redshift range $z \lesssim 0.9$ over which we have GC system photometry so far (Fig.~\ref{fig:cmd5}), to first order there is no major change in the flux distribution across the JWST/NIRCam SWC. Here, the (F115W, F150W, F200W) broadband filters, in addition to F277W from the Long Wavelength Channel (LWC), continue to be effective choices for planning observations. It is only the regime $z \gtrsim 1$ where the SED starts shifting more rapidly to longer $\lambda$ and the LWC filters will become more important.  The fact that SWC and LWC exposures can be taken simultaneously is an additional major advantage in accumulating multiwavelength sampling.

More quantitatively, the wavelength $\lambda_{peak}$ at which $L_{\nu}$ reaches a maximum is determined primarily by the $1.6\mu$ bump from the $H-$ ion \citep{sawicki2002}, which gives $\lambda_{peak} = (1+z) 1.6\mu$.  Thus for $z < 1$, we have $\lambda_{\rm peak} < 3.3$ microns; and for $z < 0.5$ (the range over which most of the lensing-cluster imaging with JWST has been done to date) we have $\lambda_{\rm peak} < 2.5$ microns, well covered by the SWC camera. The SWC also has the advantage of narrower PSFs and higher spatial resolution and pixel scale, important for minimizing effects of crowding on the photometry.

\section{Summary and Conclusions}\label{sec:summary}

Deep \jwst NIRCam imaging is used to investigate the globular cluster population in the lensing cluster MACS0417.5-1154 at $z = 0.443$. The distribution of the GC candidates in the various color-magnitude diagrams generated from the NIRCam filters (F090W, F115W, F150W, F200W) is shown, followed by a discussion of their luminosity function and their spatial distribution around the central BCG of the cluster. For the first time, the results are placed in context by direct comparison with four other BCG targets, covering a range of more than 7 Gyr in lookback time. Finally, we assess the relative merits of the NIRCam filters and cameras for imaging of GC systems over the range of redshifts that JWST has now made accessible.

A brief summary of our findings is as follows:
\begin{enumerate}
    \item[(1)] For MACS0417, the photometric limit (50\% recovery rate) depends on both magnitude and local sky noise, as found in our previous study of Abell 2744, but is near $F150W_{AB} = 29.7$ for regions of relatively low sky noise (``Zone 1''). 
    \item[(2)] The GC candidates define clear vertical sequences in the various color-magnitude diagrams constructed from the photometry, extending over 3 magnitudes before being terminated by the limiting depth of the measurements. The CMD in magnitude versus (F150W-F200W), which is relatively insensitive to GC metallicity, defines the narrowest sequence. 
    \item[(3)]  The completeness-corrected luminosity function is well matched by a simple lognormal curve with dispersion $\sigma \simeq 1.6$ magnitudes, though the current data fall well short (by perhaps 3 mag) of the expected peak or turnover point. By comparison, in Abell 2744, over a similar luminosity range the LF is steeper, with $\sigma \simeq 1.4$ mag. The high-luminosity, high-mass end of the LF is likely to be populated by a mixture of super-luminous GCs and UCDs, as has been found in other massive galaxy clusters, but the extreme upper end of the entire sequence appears to fall no higher than $10^8 M_{\odot}$. The total GC population for MACS0417 is estimated to be $N_t = 1.5 \times 10^5$, though with an uncertainty of as much as a factor of two. 
    \item[(4)] The spatial distribution of the GCs in MACS0417 centered on its BCG matches well in ellipticity and orientation with the extended halo light of the BCG. 
    \item[(5)] MACS0417 is placed in context with four other galaxy clusters with rich GC systems at redshifts extending from 0 to 0.78, covering more than half the age of the Universe in lookback time. An immediately obvious feature of the sequence of CMDs is the systematic change of the GCs due to simple stellar evolution and progressive age fading. For the first time, we are able to see with direct observational evidence the expected trends from theoretical simulations.
    \item[(6)] Lastly, we assess the utility of the various \jwst NIRCam broadband filters for further deep imaging of GC systems at redshifts up to $z = 2$. Comparing the redshifted SEDs of GCs at various redshifts with the wavelength coverage of the filters, we find that the filters in the Short Wavelength Channel are very effective choices for any targets within $z \lesssim 1$. It is only for redshifts much larger than $z \sim 1$ that the filters in the Long Wavelength Channel will become more important for deep color/magnitude surveys.
    
\end{enumerate}

\begin{acknowledgments}

The authors acknowledge support provided by Compute Ontario (\href{https://www.computeontario.ca}{https://www.computeontario.ca}) and Digital Research Alliance of Canada (\href{alliancecan.ca}{alliancecan.ca}). 
MRC gratefully acknowledges the Canadian Institute for Theoretical Astrophysics (CITA) Fellowship for support. This work was supported by the Natural Sciences and Engineering Research Council of Canada (NSERC).  MB, GR, and NM acknowledge support from the ERC Grant FIRSTLIGHT and the Slovenian national research agency ARIS through grants N1-0238 and P1-0188. This research was also enabled by grant 18JWST-GTO1 from the Canadian Space Agency.

\end{acknowledgments}

\section*{Data Availability} 
 The data presented in this article were obtained from the Mikulski Archive for Space Telescopes (MAST) at the Space Telescope Science Institute. The specific observations analyzed can be accessed via JWST images of MACS0417 at \dataset[doi:10.17909/jtaf-tr33]{https://doi.org/10.17909/jtaf-tr33}.
 
\vspace{5mm}
\facilities{\jwst (NIRCAM)}


\software{Daophot \citep{stetson1987},
          IRAF \citep{tody1986,tody1993}
          Jupyter Notebooks \citep{kluyver+2016}, 
          Matplotlib \citep{hunter2007},
          Numpy \citep{harris+2020b},
          PARSECv1.2 \citep{bressan+2012}
          }
          




\bibliography{main}{}

\begin{thebibliography}{}
\expandafter\ifx\csname natexlab\endcsname\relax\def\natexlab#1{#1}\fi
\providecommand{\url}[1]{\href{#1}{#1}}
\providecommand{\dodoi}[1]{doi:~\href{http://doi.org/#1}{\nolinkurl{#1}}}
\providecommand{\doeprint}[1]{\href{http://ascl.net/#1}{\nolinkurl{http://ascl.net/#1}}}
\providecommand{\doarXiv}[1]{\href{https://arxiv.org/abs/#1}{\nolinkurl{https://arxiv.org/abs/#1}}}

\bibitem[{A. {Adamo} {et~al.}(2024){Adamo}, {Bradley}, {Vanzella}, {Claeyssens}, {Welch}, {Diego}, {Mahler}, {Oguri}, {Sharon}, {Abdurro'uf}, {Hsiao}, {Xu}, {Messa}, {Lassen}, {Zackrisson}, {Brammer}, {Coe}, {Kokorev}, {Ricotti}, {Zitrin}, {Fujimoto}, {Inoue}, {Resseguier}, {Rigby}, {Jim{\'e}nez-Teja}, {Windhorst}, {Hashimoto}, \& {Tamura}}]{adamo+2024}
{Adamo}, A., {Bradley}, L.~D., {Vanzella}, E., {et~al.} 2024, \bibinfo{title}{{Bound star clusters observed in a lensed galaxy 460 Myr after the Big Bang},} \nat, 632, 513, \dodoi{10.1038/s41586-024-07703-7}

\bibitem[{D.~E. {Applegate} {et~al.}(2014){Applegate}, {von der Linden}, {Kelly}, {Allen}, {Allen}, {Burchat}, {Burke}, {Ebeling}, {Mantz}, \& {Morris}}]{applegate+2014}
{Applegate}, D.~E., {von der Linden}, A., {Kelly}, P.~L., {et~al.} 2014, \bibinfo{title}{{Weighing the Giants - III. Methods and measurements of accurate galaxy cluster weak-lensing masses},} \mnras, 439, 48, \dodoi{10.1093/mnras/stt2129}

\bibitem[{Y. {Asada} {et~al.}(2023){Asada}, {Sawicki}, {Desprez}, {Abraham}, {Brada{\v{c}}}, {Brammer}, {Harshan}, {Iyer}, {Martis}, {Mowla}, {Muzzin}, {Noirot}, {Ravindranath}, {Sarrouh}, {Strait}, {Willott}, \& {Zabl}}]{asada+2023}
{Asada}, Y., {Sawicki}, M., {Desprez}, G., {et~al.} 2023, \bibinfo{title}{{JWST catches the assembly of a z 5 ultra-low-mass galaxy},} \mnras, 523, L40, \dodoi{10.1093/mnrasl/slad054}

\bibitem[{Y. {Asada} {et~al.}(2024){Asada}, {Sawicki}, {Abraham}, {Brada{\v{c}}}, {Brammer}, {Desprez}, {Estrada-Carpenter}, {Iyer}, {Martis}, {Matharu}, {Mowla}, {Muzzin}, {Noirot}, {Sarrouh}, {Strait}, {Willott}, \& {Harshan}}]{asada+2024}
{Asada}, Y., {Sawicki}, M., {Abraham}, R., {et~al.} 2024, \bibinfo{title}{{Bursty star formation and galaxy-galaxy interactions in low-mass galaxies 1 Gyr after the Big Bang},} \mnras, 527, 11372, \dodoi{10.1093/mnras/stad3902}

\bibitem[{S. {Bellstedt} {et~al.}(2016){Bellstedt}, {Lidman}, {Muzzin}, {Franx}, {Guatelli}, {Hill}, {Hoekstra}, {Kurinsky}, {Labbe}, {Marchesini}, {Marsan}, {Safavi-Naeini}, {Sif{\'o}n}, {Stefanon}, {van de Sande}, {van Dokkum}, \& {Weigel}}]{bellstedt+2016}
{Bellstedt}, S., {Lidman}, C., {Muzzin}, A., {et~al.} 2016, \bibinfo{title}{{The evolution in the stellar mass of brightest cluster galaxies over the past 10 billion years},} \mnras, 460, 2862, \dodoi{10.1093/mnras/stw1184}

\bibitem[{A. {Bressan} {et~al.}(2012){Bressan}, {Marigo}, {Girardi}, {Salasnich}, {Dal Cero}, {Rubele}, \& {Nanni}}]{bressan+2012}
{Bressan}, A., {Marigo}, P., {Girardi}, L., {et~al.} 2012, \bibinfo{title}{{PARSEC: stellar tracks and isochrones with the PAdova and TRieste Stellar Evolution Code},} \mnras, 427, 127, \dodoi{10.1111/j.1365-2966.2012.21948.x}

\bibitem[{Y. {Chen} \& O.~Y. {Gnedin}(2024){Chen} \& {Gnedin}}]{chen_gnedin2024}
{Chen}, Y., \& {Gnedin}, O.~Y. 2024, \bibinfo{title}{{Galaxy assembly revealed by globular clusters},} arXiv e-prints, arXiv:2401.17420, \dodoi{10.48550/arXiv.2401.17420}

\bibitem[{K. {Chiboucas} {et~al.}(2011){Chiboucas}, {Tully}, {Marzke}, {Phillipps}, {Price}, {Peng}, {Trentham}, {Carter}, \& {Hammer}}]{chiboucas+2011}
{Chiboucas}, K., {Tully}, R.~B., {Marzke}, R.~O., {et~al.} 2011, \bibinfo{title}{{Ultra-compact Dwarfs in the Coma Cluster},} \apj, 737, 86, \dodoi{10.1088/0004-637X/737/2/86}

\bibitem[{H. {Cho} {et~al.}(2016){Cho}, {Blakeslee}, {Chies-Santos}, {Jee}, {Jensen}, {Peng}, \& {Lee}}]{cho+2016}
{Cho}, H., {Blakeslee}, J.~P., {Chies-Santos}, A.~L., {et~al.} 2016, \bibinfo{title}{{The Globular Cluster System of the Coma cD Galaxy NGC 4874 from Hubble Space Telescope ACS and WFC3/IR Imaging},} \apj, 822, 95, \dodoi{10.3847/0004-637X/822/2/95}

\bibitem[{N. {Choksi} \& O.~Y. {Gnedin}(2019){Choksi} \& {Gnedin}}]{choksi_gnedin2019}
{Choksi}, N., \& {Gnedin}, O.~Y. 2019, \bibinfo{title}{{Origins of scaling relations of globular cluster systems},} \mnras, 488, 5409, \dodoi{10.1093/mnras/stz2097}

\bibitem[{J.~J. {Condon} \& A.~M. {Matthews}(2018){Condon} \& {Matthews}}]{condon_matthews2018}
{Condon}, J.~J., \& {Matthews}, A.~M. 2018, \bibinfo{title}{{{\ensuremath{\Lambda}}CDM Cosmology for Astronomers},} \pasp, 130, 073001, \dodoi{10.1088/1538-3873/aac1b2}

\bibitem[{V. {Dornan} \& W.~E. {Harris}(2023){Dornan} \& {Harris}}]{dornan_harris2023}
{Dornan}, V., \& {Harris}, W.~E. 2023, \bibinfo{title}{{Investigating the M $_{GCS}$-M $_{ h }$ Relation in the Most Massive Galaxies},} \apj, 950, 179, \dodoi{10.3847/1538-4357/accbc3}

\bibitem[{G.~M. {Eadie} {et~al.}(2022){Eadie}, {Harris}, \& {Springford}}]{eadie+2022}
{Eadie}, G.~M., {Harris}, W.~E., \& {Springford}, A. 2022, \bibinfo{title}{{Clearing the Hurdle: The Mass of Globular Cluster Systems as a Function of Host Galaxy Mass},} \apj, 926, 162, \dodoi{10.3847/1538-4357/ac33b0}

\bibitem[{V. {Estrada-Carpenter} {et~al.}(2024){Estrada-Carpenter}, {Sawicki}, {Brammer}, {Desprez}, {Abraham}, {Asada}, {Brada{\v{c}}}, {Iyer}, {Martis}, {Matharu}, {Mowla}, {Muzzin}, {Noirot}, {Sarrouh}, {Strait}, \& {Willott}}]{estrada-carpenter+2024}
{Estrada-Carpenter}, V., {Sawicki}, M., {Brammer}, G., {et~al.} 2024, \bibinfo{title}{{When, where, and how star formation happens in a galaxy pair at cosmic noon using CANUCS JWST/NIRISS grism spectroscopy},} \mnras, 532, 577, \dodoi{10.1093/mnras/stae1368}

\bibitem[{D.~A. {Forbes}(2020){Forbes}}]{forbes2020}
{Forbes}, D.~A. 2020, \bibinfo{title}{{Reverse engineering the Milky Way},} \mnras, 493, 847, \dodoi{10.1093/mnras/staa245}

\bibitem[{D.~A. {Forbes} {et~al.}(2014){Forbes}, {Norris}, {Strader}, {Romanowsky}, {Pota}, {Kannappan}, {Brodie}, \& {Huxor}}]{forbes+2014}
{Forbes}, D.~A., {Norris}, M.~A., {Strader}, J., {et~al.} 2014, \bibinfo{title}{{The AIMSS Project II: dynamical-to-stellar mass ratios across the star cluster-galaxy divide},} \mnras, 444, 2993, \dodoi{10.1093/mnras/stu1631}

\bibitem[{D.~A. {Forbes} \& R.-S. {Remus}(2018){Forbes} \& {Remus}}]{forbes_remus2018}
{Forbes}, D.~A., \& {Remus}, R.-S. 2018, \bibinfo{title}{{Metallicity gradients in the globular cluster systems of early-type galaxies: in situ and accreted components?},} \mnras, 479, 4760, \dodoi{10.1093/mnras/sty1767}

\bibitem[{D. {Geisler} {et~al.}(1996){Geisler}, {Lee}, \& {Kim}}]{geisler+1996}
{Geisler}, D., {Lee}, M.~G., \& {Kim}, E. 1996, \bibinfo{title}{{Washington Photometry of the Globular Cluster System of NGC 4472.I.Analysis of the Metallicities},} \aj, 111, 1529, \dodoi{10.1086/117894}

\bibitem[{D.~A. {Hanes} \& D.~G. {Whittaker}(1987){Hanes} \& {Whittaker}}]{hanes_whittaker1987}
{Hanes}, D.~A., \& {Whittaker}, D.~G. 1987, \bibinfo{title}{{Globular Clusters as Extragalactic Distance Indicators: Maximum-Liklihood Methods},} \aj, 94, 906, \dodoi{10.1086/114525}

\bibitem[{C.~R. Harris {et~al.}(2020)Harris, Millman, van~der Walt, Gommers, Virtanen, Cournapeau, Wieser, Taylor, Berg, Smith, Kern, Picus, Hoyer, van Kerkwijk, Brett, Haldane, del R{\'{i}}o, Wiebe, Peterson, G{\'{e}}rard-Marchant, Sheppard, Reddy, Weckesser, Abbasi, Gohlke, \& Oliphant}]{harris+2020b}
Harris, C.~R., Millman, K.~J., van~der Walt, S.~J., {et~al.} 2020, \bibinfo{title}{Array programming with {NumPy},} Nature, 585, 357, \dodoi{10.1038/s41586-020-2649-2}

\bibitem[{W.~E. {Harris}(2009){Harris}}]{harris2009}
{Harris}, W.~E. 2009, \bibinfo{title}{{Globular Cluster Systems in Giant Ellipticals: The Mass/Metallicity Relation},} \apj, 699, 254, \dodoi{10.1088/0004-637X/699/1/254}

\bibitem[{W.~E. {Harris}(2023){Harris}}]{harris2023}
{Harris}, W.~E. 2023, \bibinfo{title}{{A Photometric Survey of Globular Cluster Systems in Brightest Cluster Galaxies},} \apjs, 265, 9, \dodoi{10.3847/1538-4365/acab5c}

\bibitem[{W.~E. {Harris} {et~al.}(2016){Harris}, {Blakeslee}, {Whitmore}, {Gnedin}, {Geisler}, \& {Rothberg}}]{harris+2016}
{Harris}, W.~E., {Blakeslee}, J.~P., {Whitmore}, B.~C., {et~al.} 2016, \bibinfo{title}{{Globular Cluster Systems in Brightest Cluster Galaxies. II. NGC 6166},} \apj, 817, 58, \dodoi{10.3847/0004-637X/817/1/58}

\bibitem[{W.~E. {Harris} \& M. {Reina-Campos}(2023){Harris} \& {Reina-Campos}}]{harris_reinacampos2023}
{Harris}, W.~E., \& {Reina-Campos}, M. 2023, \bibinfo{title}{{JWST photometry of globular cluster populations in Abell 2744 at z = 0.3},} \mnras, 526, 2696, \dodoi{10.1093/mnras/stad2903}

\bibitem[{W.~E. {Harris} \& M. {Reina-Campos}(2024){Harris} \& {Reina-Campos}}]{harris_reinacampos2024}
{Harris}, W.~E., \& {Reina-Campos}, M. 2024, \bibinfo{title}{{JWST Photometry of Globular Clusters in Abell 2744. II: luminosity and color distributions},} arXiv e-prints, arXiv:2404.10813, \dodoi{10.48550/arXiv.2404.10813}

\bibitem[{W.~E. {Harris} \& J. {Speagle}(2024){Harris} \& {Speagle}}]{harris_speagle2024}
{Harris}, W.~E., \& {Speagle}, J. 2024, \bibinfo{title}{{Photometric Completeness Modelled with a Neural Network},} submitted

\bibitem[{W.~E. {Harris} {et~al.}(2014){Harris}, {Morningstar}, {Gnedin}, {O'Halloran}, {Blakeslee}, {Whitmore}, {C{\^o}t{\'e}}, {Geisler}, {Peng}, {Bailin}, {Rothberg}, {Cockcroft}, \& {Barber DeGraaff}}]{harris+2014}
{Harris}, W.~E., {Morningstar}, W., {Gnedin}, O.~Y., {et~al.} 2014, \bibinfo{title}{{Globular Cluster Systems in Brightest Cluster Galaxies: A Near-universal Luminosity Function?},} \apj, 797, 128, \dodoi{10.1088/0004-637X/797/2/128}

\bibitem[{W.~E. {Harris} {et~al.}(2025){Harris}, {Reina-Campos}, {Koekemoer}, {Berkheimer}, {Carleton}, {Cohen}, {Frye}, {Hinrichs}, {Holwerda}, {Honor}, {Ricotti}, {Willner}, {Windhorst}, \& {Yan}}]{harris+2025}
{Harris}, W.~E., {Reina-Campos}, M., {Koekemoer}, A.~M., {et~al.} 2025, \bibinfo{title}{{PEARLS: Globular Clusters and Ultra-Compact Dwarfs in the El Gordo Galaxies at z=0.87},} arXiv e-prints, arXiv:2508.12862.
\newblock \doarXiv{2508.12862}

\bibitem[{K. {Hartman} {et~al.}(2023){Hartman}, {Harris}, {Blakeslee}, {Ma}, \& {Greene}}]{hartman+2023}
{Hartman}, K., {Harris}, W.~E., {Blakeslee}, J.~P., {Ma}, C.-P., \& {Greene}, J.~E. 2023, \bibinfo{title}{{Comparing Globular Cluster System Properties with Host Galaxy Environment},} \apj, 953, 154, \dodoi{10.3847/1538-4357/ace340}

\bibitem[{G.~K.~T. {Hau} {et~al.}(2009){Hau}, {Spitler}, {Forbes}, {Proctor}, {Strader}, {Mendel}, {Brodie}, \& {Harris}}]{hau+2009}
{Hau}, G. K.~T., {Spitler}, L.~R., {Forbes}, D.~A., {et~al.} 2009, \bibinfo{title}{{An ultra-compact dwarf around the Sombrero galaxy (M104): the nearest massive UCD},} \mnras, 394, L97, \dodoi{10.1111/j.1745-3933.2009.00618.x}

\bibitem[{M. {Hilker} {et~al.}(2007){Hilker}, {Baumgardt}, {Infante}, {Drinkwater}, {Evstigneeva}, \& {Gregg}}]{hilker+2007}
{Hilker}, M., {Baumgardt}, H., {Infante}, L., {et~al.} 2007, \bibinfo{title}{{Dynamical masses of ultra-compact dwarf galaxies in Fornax},} \aap, 463, 119, \dodoi{10.1051/0004-6361:20066429}

\bibitem[{E. {Hubble} \& R.~C. {Tolman}(1935){Hubble} \& {Tolman}}]{hubble_tolman1935}
{Hubble}, E., \& {Tolman}, R.~C. 1935, \bibinfo{title}{{Two Methods of Investigating the Nature of the Nebular Redshift},} \apj, 82, 302, \dodoi{10.1086/143682}

\bibitem[{J.~D. {Hunter}(2007){Hunter}}]{hunter2007}
{Hunter}, J.~D. 2007, \bibinfo{title}{{Matplotlib: A 2D Graphics Environment},} Computing in Science and Engineering, 9, 90, \dodoi{10.1109/MCSE.2007.55}

\bibitem[{S. {Janssens} {et~al.}(2017){Janssens}, {Abraham}, {Brodie}, {Forbes}, {Romanowsky}, \& {van Dokkum}}]{janssens+2017}
{Janssens}, S., {Abraham}, R., {Brodie}, J., {et~al.} 2017, \bibinfo{title}{{Ultra-diffuse and Ultra-compact Galaxies in the Frontier Fields Cluster Abell 2744},} \apjl, 839, L17, \dodoi{10.3847/2041-8213/aa667d}

\bibitem[{S.~R. {Janssens} {et~al.}(2019){Janssens}, {Abraham}, {Brodie}, {Forbes}, \& {Romanowsky}}]{janssens+2019}
{Janssens}, S.~R., {Abraham}, R., {Brodie}, J., {Forbes}, D.~A., \& {Romanowsky}, A.~J. 2019, \bibinfo{title}{{The Distribution of Ultra-diffuse and Ultra-compact Galaxies in the Frontier Fields},} \apj, 887, 92, \dodoi{10.3847/1538-4357/ab536c}

\bibitem[{J. {Janz} {et~al.}(2015){Janz}, {Forbes}, {Norris}, {Strader}, {Penny}, {Fagioli}, \& {Romanowsky}}]{janz+2015}
{Janz}, J., {Forbes}, D.~A., {Norris}, M.~A., {et~al.} 2015, \bibinfo{title}{{How elevated is the dynamical-to-stellar mass ratio of the ultracompact dwarf S999?},} \mnras, 449, 1716, \dodoi{10.1093/mnras/stv389}

\bibitem[{M. {Jauzac} {et~al.}(2019){Jauzac}, {Mahler}, {Edge}, {Sharon}, {Gillman}, {Ebeling}, {Harvey}, {Richard}, {Hamer}, {Fumagalli}, {Mark Swinbank}, {Kneib}, {Massey}, \& {Salom{\'e}}}]{jauzac+2019}
{Jauzac}, M., {Mahler}, G., {Edge}, A.~C., {et~al.} 2019, \bibinfo{title}{{The core of the massive cluster merger MACS J0417.5-1154 as seen by VLT/MUSE},} \mnras, 483, 3082, \dodoi{10.1093/mnras/sty3312}

\bibitem[{P.~S. {Joschko} {et~al.}(2024){Joschko}, {Kruijssen}, {Trujillo-Gomez}, {Pfeffer}, {Bastian}, {Crain}, \& {Reina-Campos}}]{joschko+2024}
{Joschko}, P.~S., {Kruijssen}, J.~M.~D., {Trujillo-Gomez}, S., {et~al.} 2024, \bibinfo{title}{{The cosmic globular cluster formation history in the E-MOSAICS simulations},} arXiv e-prints, arXiv:2412.04105, \dodoi{10.48550/arXiv.2412.04105}

\bibitem[{K.~E. {Keatley} \& W.~E. {Harris}(2025){Keatley} \& {Harris}}]{keatley_harris2025}
{Keatley}, K.~E., \& {Harris}, W.~E. 2025, \bibinfo{title}{{JWST NIRCam Observations of the Globular Cluster Population of RXJ 2129.7+0005},} arXiv e-prints, arXiv:2508.03626, \dodoi{10.48550/arXiv.2508.03626}

\bibitem[{T. {Kluyver} {et~al.}(2016){Kluyver}, {Ragan-Kelley}, Fernando, {Granger}, {Bussonnier}, {Frederic}, {Kelley}, {Hamrick}, {Grout}, {Corlay}, {Ivanov}, {Avila}, {Abdalla}, {Willing}, \& {Jupyter Development Team}}]{kluyver+2016}
{Kluyver}, T., {Ragan-Kelley}, B., Fernando, P., {et~al.} 2016, {Jupyter Notebooks - a publishing format for reproducible computational workflows}, 87--90, \dodoi{10.3233/978-1-61499-649-1-87}

\bibitem[{J.~M.~D. {Kruijssen} {et~al.}(2020){Kruijssen}, {Pfeffer}, {Chevance}, {Bonaca}, {Trujillo-Gomez}, {Bastian}, {Reina-Campos}, {Crain}, \& {Hughes}}]{kruijssen+2020}
{Kruijssen}, J.~M.~D., {Pfeffer}, J.~L., {Chevance}, M., {et~al.} 2020, \bibinfo{title}{{Kraken reveals itself - the merger history of the Milky Way reconstructed with the E-MOSAICS simulations},} \mnras, 498, 2472, \dodoi{10.1093/mnras/staa2452}

\bibitem[{R. {Leaman} {et~al.}(2013){Leaman}, {VandenBerg}, \& {Mendel}}]{leaman_vandenberg2013}
{Leaman}, R., {VandenBerg}, D.~A., \& {Mendel}, J.~T. 2013, \bibinfo{title}{{The bifurcated age-metallicity relation of Milky Way globular clusters and its implications for the accretion history of the galaxy},} \mnras, 436, 122, \dodoi{10.1093/mnras/stt1540}

\bibitem[{H. {Li} \& O.~Y. {Gnedin}(2019){Li} \& {Gnedin}}]{li_gnedin2019}
{Li}, H., \& {Gnedin}, O.~Y. 2019, \bibinfo{title}{{Star cluster formation in cosmological simulations - III. Dynamical and chemical evolution},} \mnras, 486, 4030, \dodoi{10.1093/mnras/stz1114}

\bibitem[{C. {Liu} {et~al.}(2011){Liu}, {Peng}, {Jord{\'a}n}, {Ferrarese}, {Blakeslee}, {C{\^o}t{\'e}}, \& {Mei}}]{liu+2011}
{Liu}, C., {Peng}, E.~W., {Jord{\'a}n}, A., {et~al.} 2011, \bibinfo{title}{{The ACS Fornax Cluster Survey. X. Color Gradients of Globular Cluster Systems in Early-type Galaxies},} \apj, 728, 116, \dodoi{10.1088/0004-637X/728/2/116}

\bibitem[{C. {Liu} {et~al.}(2020){Liu}, {C{\^o}t{\'e}}, {Peng}, {Roediger}, {Zhang}, {Ferrarese}, {S{\'a}nchez-Janssen}, {Guhathakurta}, {Yang}, {Jing}, {Alamo-Mart{\'\i}nez}, {Blakeslee}, {Boselli}, {Cuilandre}, {Duc}, {Durrell}, {Gwyn}, {Jord{\'a}n}, {Ko}, {Lan{\c{c}}on}, {Lim}, {Longobardi}, {Mei}, {Mihos}, {Mu{\~n}oz}, {Powalka}, {Puzia}, {Spengler}, \& {Toloba}}]{liu+2020}
{Liu}, C., {C{\^o}t{\'e}}, P., {Peng}, E.~W., {et~al.} 2020, \bibinfo{title}{{The Next Generation Virgo Cluster Survey. XXXIV. Ultracompact Dwarf Galaxies in the Virgo Cluster},} \apjs, 250, 17, \dodoi{10.3847/1538-4365/abad91}

\bibitem[{L.~M. {Lubin} \& A. {Sandage}(2001){Lubin} \& {Sandage}}]{lubin_sandage2001}
{Lubin}, L.~M., \& {Sandage}, A. 2001, \bibinfo{title}{{The Tolman Surface Brightness Test for the Reality of the Expansion. IV. A Measurement of the Tolman Signal and the Luminosity Evolution of Early-Type Galaxies},} \aj, 122, 1084, \dodoi{10.1086/322134}

\bibitem[{G. {Mahler} {et~al.}(2019){Mahler}, {Sharon}, {Fox}, {Coe}, {Jauzac}, {Strait}, {Edge}, {Acebron}, {Andrade-Santos}, {Avila}, {Brada{\v{c}}}, {Bradley}, {Carrasco}, {Cerny}, {Cibirka}, {Czakon}, {Dawson}, {Frye}, {Hoag}, {Huang}, {Johnson}, {Jones}, {Kikuchihara}, {Lam}, {Livermore}, {Lovisari}, {Mainali}, {Ogaz}, {Ouchi}, {Paterno-Mahler}, {Roederer}, {Ryan}, {Salmon}, {Sendra-Server}, {Stark}, {Toft}, {Trenti}, {Umetsu}, {Vulcani}, \& {Zitrin}}]{mahler+2019}
{Mahler}, G., {Sharon}, K., {Fox}, C., {et~al.} 2019, \bibinfo{title}{{RELICS: Strong Lensing Analysis of MACS J0417.5-1154 and Predictions for Observing the Magnified High-redshift Universe with JWST},} \apj, 873, 96, \dodoi{10.3847/1538-4357/ab042b}

\bibitem[{A.~B. {Mantz} {et~al.}(2015){Mantz}, {Allen}, {Morris}, {Schmidt}, {von der Linden}, \& {Urban}}]{mantz+2015}
{Mantz}, A.~B., {Allen}, S.~W., {Morris}, R.~G., {et~al.} 2015, \bibinfo{title}{{Cosmology and astrophysics from relaxed galaxy clusters - I. Sample selection},} \mnras, 449, 199, \dodoi{10.1093/mnras/stv219}

\bibitem[{D.~E. {McLaughlin} {et~al.}(1994){McLaughlin}, {Harris}, \& {Hanes}}]{mclaughlin+1994}
{McLaughlin}, D.~E., {Harris}, W.~E., \& {Hanes}, D.~A. 1994, \bibinfo{title}{{The Spatial Structure of the M87 Globular Cluster System},} \apj, 422, 486, \dodoi{10.1086/173744}

\bibitem[{M. {Messa} {et~al.}(2024){Messa}, {Dessauges-Zavadsky}, {Adamo}, {Richard}, \& {Claeyssens}}]{fujimoto+2024}
{Messa}, M., {Dessauges-Zavadsky}, M., {Adamo}, A., {Richard}, J., \& {Claeyssens}, A. 2024, \bibinfo{title}{{Properties of the brightest young stellar clumps in extremely lensed galaxies at redshifts 4 to 5},} \mnras, 529, 2162, \dodoi{10.1093/mnras/stae565}

\bibitem[{L. {Mowla} {et~al.}(2022){Mowla}, {Iyer}, {Desprez}, {Estrada-Carpenter}, {Martis}, {Noirot}, {Sarrouh}, {Strait}, {Asada}, {Abraham}, {Brammer}, {Sawicki}, {Willott}, {Bradac}, {Doyon}, {Muzzin}, {Pacifici}, {Ravindranath}, \& {Zabl}}]{mowla+2022}
{Mowla}, L., {Iyer}, K.~G., {Desprez}, G., {et~al.} 2022, \bibinfo{title}{{The Sparkler: Evolved High-redshift Globular Cluster Candidates Captured by JWST},} \apjl, 937, L35, \dodoi{10.3847/2041-8213/ac90ca}

\bibitem[{L. {Mowla} {et~al.}(2024){Mowla}, {Iyer}, {Asada}, {Desprez}, {Tan}, {Martis}, {Sarrouh}, {Strait}, {Abraham}, {Brada{\v{c}}}, {Brammer}, {Muzzin}, {Pacifici}, {Ravindranath}, {Sawicki}, {Willott}, {Estrada-Carpenter}, {Jahan}, {Noirot}, {Matharu}, {Rihtar{\v{s}}i{\v{c}}}, \& {Zabl}}]{mowla+2024}
{Mowla}, L., {Iyer}, K., {Asada}, Y., {et~al.} 2024, \bibinfo{title}{{Formation of a low-mass galaxy from star clusters in a 600-million-year-old Universe},} \nat, 636, 332, \dodoi{10.1038/s41586-024-08293-0}

\bibitem[{A.~L. {Muratov} \& O.~Y. {Gnedin}(2010){Muratov} \& {Gnedin}}]{muratov_gnedin2010}
{Muratov}, A.~L., \& {Gnedin}, O.~Y. 2010, \bibinfo{title}{{Modeling the Metallicity Distribution of Globular Clusters},} \apj, 718, 1266, \dodoi{10.1088/0004-637X/718/2/1266}

\bibitem[{M.~A. {Norris} \& S.~J. {Kannappan}(2011){Norris} \& {Kannappan}}]{norris+2011}
{Norris}, M.~A., \& {Kannappan}, S.~J. 2011, \bibinfo{title}{{The ubiquity and dual nature of ultra-compact dwarfs},} \mnras, 414, 739, \dodoi{10.1111/j.1365-2966.2011.18440.x}

\bibitem[{T. {Okabe} {et~al.}(2020){Okabe}, {Oguri}, {Peirani}, {Suto}, {Dubois}, {Pichon}, {Kitayama}, {Sasaki}, \& {Nishimichi}}]{okabe+2020}
{Okabe}, T., {Oguri}, M., {Peirani}, S., {et~al.} 2020, \bibinfo{title}{{Shapes and alignments of dark matter haloes and their brightest cluster galaxies in 39 strong lensing clusters},} \mnras, 496, 2591, \dodoi{10.1093/mnras/staa1479}

\bibitem[{M.~B. {Pandge} {et~al.}(2019){Pandge}, {Monteiro-Oliveira}, {Bagchi}, {Simionescu}, {Limousin}, \& {Raychaudhury}}]{pandge+2019}
{Pandge}, M.~B., {Monteiro-Oliveira}, R., {Bagchi}, J., {et~al.} 2019, \bibinfo{title}{{A combined X-ray, optical, and radio view of the merging galaxy cluster MACS J0417.5-1154},} \mnras, 482, 5093, \dodoi{10.1093/mnras/sty2937}

\bibitem[{E.~W. {Peng} {et~al.}(2006){Peng}, {Jord{\'a}n}, {C{\^o}t{\'e}}, {Blakeslee}, {Ferrarese}, {Mei}, {West}, {Merritt}, {Milosavljevi{\'c}}, \& {Tonry}}]{peng+2006}
{Peng}, E.~W., {Jord{\'a}n}, A., {C{\^o}t{\'e}}, P., {et~al.} 2006, \bibinfo{title}{{The ACS Virgo Cluster Survey. IX. The Color Distributions of Globular Cluster Systems in Early-Type Galaxies},} \apj, 639, 95, \dodoi{10.1086/498210}

\bibitem[{S.~J. {Penny} {et~al.}(2012){Penny}, {Forbes}, \& {Conselice}}]{penny+2012}
{Penny}, S.~J., {Forbes}, D.~A., \& {Conselice}, C.~J. 2012, \bibinfo{title}{{Hubble Space Telescope survey of the Perseus cluster - IV. Compact stellar systems in the Perseus cluster core and ultracompact dwarf formation in star-forming filaments},} \mnras, 422, 885, \dodoi{10.1111/j.1365-2966.2012.20669.x}

\bibitem[{J. {Pfeffer} {et~al.}(2025){Pfeffer}, {Forbes}, {Romanowsky}, {Bastian}, {Crain}, {Kruijssen}, {Bekki}, {Brodie}, {Chevance}, {Couch}, \& {Gannon}}]{pfeffer+2025}
{Pfeffer}, J., {Forbes}, D.~A., {Romanowsky}, A.~J., {et~al.} 2025, \bibinfo{title}{{Comparing E-MOSAICS predictions of high-redshift proto-globular clusters with JWST observations in lensed galaxies},} \mnras, 536, 1878, \dodoi{10.1093/mnras/stae2665}

\bibitem[{ {Planck Collaboration} {et~al.}(2016{\natexlab{a}}){Planck Collaboration}, {Ade}, {Aghanim}, {Arnaud}, {Ashdown}, {Aumont}, {Baccigalupi}, {Banday}, {Barreiro}, {Barrena}, \& et~al.}]{planck+2016}
{Planck Collaboration}, {Ade}, P.~A.~R., {Aghanim}, N., {et~al.} 2016{\natexlab{a}}, \bibinfo{title}{{Planck 2015 results. XXVII. The second Planck catalogue of Sunyaev-Zeldovich sources},} \aap, 594, A27, \dodoi{10.1051/0004-6361/201525823}

\bibitem[{ {Planck Collaboration} {et~al.}(2016{\natexlab{b}}){Planck Collaboration}, {Ade}, {Aghanim}, {Arnaud}, {Ashdown}, {Aumont}, {Baccigalupi}, {Banday}, {Barreiro}, {Bartlett}, {Bartolo}, {Battaner}, {Battye}, {Benabed}, {Beno{\^\i}t}, {Benoit-L{\'e}vy}, {Bernard}, {Bersanelli}, {Bielewicz}, {Bock}, {Bonaldi}, {Bonavera}, {Bond}, {Borrill}, {Bouchet}, {Boulanger}, {Bucher}, {Burigana}, {Butler}, {Calabrese}, {Cardoso}, {Catalano}, {Challinor}, {Chamballu}, {Chary}, {Chiang}, {Chluba}, {Christensen}, {Church}, {Clements}, {Colombi}, {Colombo}, {Combet}, {Coulais}, {Crill}, {Curto}, {Cuttaia}, {Danese}, {Davies}, {Davis}, {de Bernardis}, {de Rosa}, {de Zotti}, {Delabrouille}, {D{\'e}sert}, {Di Valentino}, {Dickinson}, {Diego}, {Dolag}, {Dole}, {Donzelli}, {Dor{\'e}}, {Douspis}, {Ducout}, {Dunkley}, {Dupac}, {Efstathiou}, {Elsner}, {En{\ss}lin}, {Eriksen}, {Farhang}, {Fergusson}, {Finelli}, {Forni}, {Frailis}, {Fraisse}, {Franceschi}, {Frejsel}, {Galeotta}, {Galli}, {Ganga}, {Gauthier}, {Gerbino},
  {Ghosh}, {Giard}, {Giraud-H{\'e}raud}, {Giusarma}, {Gjerl{\o}w}, {Gonz{\'a}lez-Nuevo}, {G{\'o}rski}, {Gratton}, {Gregorio}, {Gruppuso}, {Gudmundsson}, {Hamann}, {Hansen}, {Hanson}, {Harrison}, {Helou}, {Henrot-Versill{\'e}}, {Hern{\'a}ndez-Monteagudo}, {Herranz}, {Hildebrandt}, {Hivon}, {Hobson}, {Holmes}, {Hornstrup}, {Hovest}, {Huang}, {Huffenberger}, {Hurier}, {Jaffe}, {Jaffe}, {Jones}, {Juvela}, {Keih{\"a}nen}, {Keskitalo}, {Kisner}, {Kneissl}, {Knoche}, {Knox}, {Kunz}, {Kurki-Suonio}, {Lagache}, {L{\"a}hteenm{\"a}ki}, {Lamarre}, {Lasenby}, {Lattanzi}, {Lawrence}, {Leahy}, {Leonardi}, {Lesgourgues}, {Levrier}, {Lewis}, {Liguori}, {Lilje}, {Linden-V{\o}rnle}, {L{\'o}pez-Caniego}, {Lubin}, {Mac{\'\i}as-P{\'e}rez}, {Maggio}, {Maino}, {Mandolesi}, {Mangilli}, {Marchini}, {Maris}, {Martin}, {Martinelli}, {Mart{\'\i}nez-Gonz{\'a}lez}, {Masi}, {Matarrese}, {McGehee}, {Meinhold}, {Melchiorri}, {Melin}, {Mendes}, {Mennella}, {Migliaccio}, {Millea}, {Mitra}, {Miville-Desch{\^e}nes}, {Moneti}, {Montier},
  {Morgante}, {Mortlock}, {Moss}, {Munshi}, {Murphy}, {Naselsky}, {Nati}, {Natoli}, {Netterfield}, {N{\o}rgaard-Nielsen}, {Noviello}, {Novikov}, {Novikov}, {Oxborrow}, {Paci}, {Pagano}, {Pajot}, {Paladini}, {Paoletti}, {Partridge}, {Pasian}, {Patanchon}, {Pearson}, {Perdereau}, {Perotto}, {Perrotta}, {Pettorino}, {Piacentini}, {Piat}, {Pierpaoli}, {Pietrobon}, {Plaszczynski}, {Pointecouteau}, {Polenta}, {Popa}, {Pratt}, {Pr{\'e}zeau}, {Prunet}, {Puget}, {Rachen}, {Reach}, {Rebolo}, {Reinecke}, {Remazeilles}, {Renault}, {Renzi}, {Ristorcelli}, {Rocha}, {Rosset}, {Rossetti}, {Roudier}, {Rouill{\'e} d'Orfeuil}, {Rowan-Robinson}, {Rubi{\~n}o-Mart{\'\i}n}, {Rusholme}, {Said}, {Salvatelli}, {Salvati}, {Sandri}, {Santos}, {Savelainen}, {Savini}, {Scott}, {Seiffert}, {Serra}, {Shellard}, {Spencer}, {Spinelli}, {Stolyarov}, {Stompor}, {Sudiwala}, {Sunyaev}, {Sutton}, {Suur-Uski}, {Sygnet}, {Tauber}, {Terenzi}, {Toffolatti}, {Tomasi}, {Tristram}, {Trombetti}, {Tucci}, {Tuovinen}, {T{\"u}rler}, {Umana}, {Valenziano},
  {Valiviita}, {Van Tent}, {Vielva}, {Villa}, {Wade}, {Wandelt}, {Wehus}, {White}, {White}, {Wilkinson}, {Yvon}, {Zacchei}, \& {Zonca}}]{planck2016}
{Planck Collaboration}, {Ade}, P.~A.~R., {Aghanim}, N., {et~al.} 2016{\natexlab{b}}, \bibinfo{title}{{Planck 2015 results. XIII. Cosmological parameters},} \aap, 594, A13, \dodoi{10.1051/0004-6361/201525830}

\bibitem[{R.~T. {Pomeroy} {et~al.}(2025){Pomeroy}, {Madrid}, {O'Neill}, \& {Gagliano}}]{pomeroy+2025}
{Pomeroy}, R.~T., {Madrid}, J.~P., {O'Neill}, C.~R., \& {Gagliano}, A.~T. 2025, \bibinfo{title}{{A Wide Field Map of Ultra-Compact Dwarfs in the Coma Cluster},} arXiv e-prints, arXiv:2506.02296, \dodoi{10.48550/arXiv.2506.02296}

\bibitem[{M. {Reina-Campos} \& W.~E. {Harris}(2023){Reina-Campos} \& {Harris}}]{reina-campos_harris2024}
{Reina-Campos}, M., \& {Harris}, W.~E. 2023, \bibinfo{title}{{RESCUER: Cosmological K-corrections for star clusters},} arXiv e-prints, arXiv:2310.02307, \dodoi{10.48550/arXiv.2310.02307}

\bibitem[{M. {Reina-Campos} {et~al.}(2022{\natexlab{a}}){Reina-Campos}, {Keller}, {Kruijssen}, {Gensior}, {Trujillo-Gomez}, {Jeffreson}, {Pfeffer}, \& {Sills}}]{reina-campos+2022b}
{Reina-Campos}, M., {Keller}, B.~W., {Kruijssen}, J.~M.~D., {et~al.} 2022{\natexlab{a}}, \bibinfo{title}{{Introducing EMP-Pathfinder: modelling the simultaneous formation and evolution of stellar clusters in their host galaxies},} \mnras, 517, 3144, \dodoi{10.1093/mnras/stac1934}

\bibitem[{M. {Reina-Campos} {et~al.}(2022{\natexlab{b}}){Reina-Campos}, {Trujillo-Gomez}, {Deason}, {Kruijssen}, {Pfeffer}, {Crain}, {Bastian}, \& {Hughes}}]{reina-campos+2022}
{Reina-Campos}, M., {Trujillo-Gomez}, S., {Deason}, A.~J., {et~al.} 2022{\natexlab{b}}, \bibinfo{title}{{Radial distributions of globular clusters trace their host dark matter halo: insights from the E-MOSAICS simulations},} \mnras, 513, 3925, \dodoi{10.1093/mnras/stac1126}

\bibitem[{B. {R{\"o}ck} {et~al.}(2016){R{\"o}ck}, {Vazdekis}, {Ricciardelli}, {Peletier}, {Knapen}, \& {Falc{\'o}n-Barroso}}]{rock+2016}
{R{\"o}ck}, B., {Vazdekis}, A., {Ricciardelli}, E., {et~al.} 2016, \bibinfo{title}{{MILES extended: Stellar population synthesis models from the optical to the infrared},} \aap, 589, A73, \dodoi{10.1051/0004-6361/201527570}

\bibitem[{E. {Rosolowsky} {et~al.}(2021){Rosolowsky}, {Hughes}, {Leroy}, {Sun}, {Querejeta}, {Schruba}, {Usero}, {Herrera}, {Liu}, {Pety}, {Saito}, {Be{\v{s}}li{\'c}}, {Bigiel}, {Blanc}, {Chevance}, {Dale}, {Deger}, {Faesi}, {Glover}, {Henshaw}, {Klessen}, {Kruijssen}, {Larson}, {Lee}, {Meidt}, {Mok}, {Schinnerer}, {Thilker}, \& {Williams}}]{rosolowsky+2021}
{Rosolowsky}, E., {Hughes}, A., {Leroy}, A.~K., {et~al.} 2021, \bibinfo{title}{{Giant molecular cloud catalogues for PHANGS-ALMA: methods and initial results},} \mnras, 502, 1218, \dodoi{10.1093/mnras/stab085}

\bibitem[{G.~T.~E. {Sarrouh} {et~al.}(2025){Sarrouh}, {Asada}, {Martis}, {Willott}, {Iyer}, {Noirot}, {Muzzin}, {Sawicki}, {Brammer}, {Desprez}, {Rihtar{\v{s}}i{\v{c}}}, {Zabl}, {Abraham}, {Brada{\v{c}}}, {Doyon}, {Antwi-Danso}, {Berek}, {Brown}, {Estrada-Carpenter}, {Favaro}, {Felicioni}, {Forrest}, {Gaspar}, {Gould}, {Gledhill}, {Harshan}, {Jahan}, {Jagga}, {Jude{\v{z}}}, {Marchesini}, {Markov}, {Matharu}, {MacFarland}, {Merchant}, {M{\'e}rida}, {Mowla}, {Myers}, {Omori}, {Pacifici}, {Ravindranath}, {Robbins}, {Strait}, {Sok}, {Tan}, {Tripodi}, {Wilson}, \& {Withers}}]{sarrouh+2025}
{Sarrouh}, G. T.~E., {Asada}, Y., {Martis}, N.~S., {et~al.} 2025, \bibinfo{title}{{CANUCS/Technicolor Data Release 1: Imaging, Photometry, Slit Spectroscopy, and Stellar Population Parameters},} arXiv e-prints, arXiv:2506.21685, \dodoi{10.48550/arXiv.2506.21685}

\bibitem[{M. {Sawicki}(2002){Sawicki}}]{sawicki2002}
{Sawicki}, M. 2002, \bibinfo{title}{{The 1.6 Micron Bump as a Photometric Redshift Indicator},} \aj, 124, 3050, \dodoi{10.1086/344682}

\bibitem[{P.~B. {Stetson}(1987){Stetson}}]{stetson1987}
{Stetson}, P.~B. 1987, \bibinfo{title}{{DAOPHOT: A Computer Program for Crowded-Field Stellar Photometry},} \pasp, 99, 191, \dodoi{10.1086/131977}

\bibitem[{D. {Tody}(1986){Tody}}]{tody1986}
{Tody}, D. 1986, in Society of Photo-Optical Instrumentation Engineers (SPIE) Conference Series, Vol. 627, Instrumentation in astronomy VI, ed. D.~L. {Crawford}, 733, \dodoi{10.1117/12.968154}

\bibitem[{D. {Tody}(1993){Tody}}]{tody1993}
{Tody}, D. 1993, in Astronomical Society of the Pacific Conference Series, Vol.~52, Astronomical Data Analysis Software and Systems II, ed. R.~J. {Hanisch}, R.~J.~V. {Brissenden}, \& J.~{Barnes}, 173

\bibitem[{C. {Usher} {et~al.}(2024){Usher}, {Caldwell}, \& {Cabrera-Ziri}}]{usher+2024}
{Usher}, C., {Caldwell}, N., \& {Cabrera-Ziri}, I. 2024, \bibinfo{title}{{Measuring M31 globular cluster ages and metallicities using both photometry and spectroscopy},} \mnras, 528, 6010, \dodoi{10.1093/mnras/stae282}

\bibitem[{D.~A. {VandenBerg} {et~al.}(2013){VandenBerg}, {Brogaard}, {Leaman}, \& {Casagrande}}]{vandenberg+2013}
{VandenBerg}, D.~A., {Brogaard}, K., {Leaman}, R., \& {Casagrande}, L. 2013, \bibinfo{title}{{The Ages of 55 Globular Clusters as Determined Using an Improved \textbackslashDelta V\^HB\_TO Method along with Color-Magnitude Diagram Constraints, and Their Implications for Broader Issues},} \apj, 775, 134, \dodoi{10.1088/0004-637X/775/2/134}

\bibitem[{E. {Vanzella} {et~al.}(2022){Vanzella}, {Castellano}, {Bergamini}, {Treu}, {Mercurio}, {Scarlata}, {Rosati}, {Grillo}, {Acebron}, {Caminha}, {Nonino}, {Nanayakkara}, {Roberts-Borsani}, {Bradac}, {Wang}, {Brammer}, {Strait}, {Vulcani}, {Me{\v{s}}tri{\'c}}, {Meneghetti}, {Calura}, {Henry}, {Zanella}, {Trenti}, {Boyett}, {Morishita}, {Calabr{\`o}}, {Glazebrook}, {Marchesini}, {Birrer}, {Yang}, \& {Jones}}]{vanzella+2022}
{Vanzella}, E., {Castellano}, M., {Bergamini}, P., {et~al.} 2022, \bibinfo{title}{{Early Results from GLASS-JWST. VII. Evidence for Lensed, Gravitationally Bound Protoglobular Clusters at z = 4 in the Hubble Frontier Field A2744},} \apjl, 940, L53, \dodoi{10.3847/2041-8213/ac8c2d}

\bibitem[{E. {Vanzella} {et~al.}(2023){Vanzella}, {Claeyssens}, {Welch}, {Adamo}, {Coe}, {Diego}, {Mahler}, {Khullar}, {Kokorev}, {Oguri}, {Ravindranath}, {Furtak}, {Hsiao}, {Abdurro'uf}, {Mandelker}, {Brammer}, {Bradley}, {Brada{\v{c}}}, {Conselice}, {Dayal}, {Nonino}, {Andrade-Santos}, {Windhorst}, {Pirzkal}, {Sharon}, {de Mink}, {Fujimoto}, {Zitrin}, {Eldridge}, \& {Norman}}]{vanzella+2023}
{Vanzella}, E., {Claeyssens}, A., {Welch}, B., {et~al.} 2023, \bibinfo{title}{{JWST/NIRCam Probes Young Star Clusters in the Reionization Era Sunrise Arc},} \apj, 945, 53, \dodoi{10.3847/1538-4357/acb59a}

\bibitem[{D. {Villegas} {et~al.}(2010){Villegas}, {Jord{\'a}n}, {Peng}, {Blakeslee}, {C{\^o}t{\'e}}, {Ferrarese}, {Kissler-Patig}, {Mei}, {Infante}, {Tonry}, \& {West}}]{villegas+2010}
{Villegas}, D., {Jord{\'a}n}, A., {Peng}, E.~W., {et~al.} 2010, \bibinfo{title}{{The ACS Fornax Cluster Survey. VIII. The Luminosity Function of Globular Clusters in Virgo and Fornax Early-type Galaxies and Its Use as a Distance Indicator},} \apj, 717, 603, \dodoi{10.1088/0004-637X/717/2/603}

\bibitem[{K.~E. {Whitaker} {et~al.}(2025){Whitaker}, {Cutler}, {Chandar}, {Pan}, {Setton}, {Furtak}, {Bezanson}, {Labb{\'e}}, {Leja}, {Suess}, {Wang}, {Weaver}, {Atek}, {Brammer}, {Feldmann}, {F{\"o}rster Schreiber}, {Glazebrook}, {de Graaff}, {Greene}, {Khullar}, {Marchesini}, {Maseda}, {Miller}, {Mo}, {Mowla}, {Nanayakkara}, {Nelson}, {Price}, {Rizzo}, {van Dokkum}, {Williams}, {Zhang}, {Zhang}, \& {Zitrin}}]{whitaker+2025}
{Whitaker}, K.~E., {Cutler}, S.~E., {Chandar}, R., {et~al.} 2025, \bibinfo{title}{{Discovery of Ancient Globular Cluster Candidates in The Relic, a Quiescent Galaxy at z=2.5},} arXiv e-prints, arXiv:2501.07627, \dodoi{10.48550/arXiv.2501.07627}

\bibitem[{C.~J. {Willott} {et~al.}(2022){Willott}, {Doyon}, {Albert}, {Brammer}, {Dixon}, {Muzic}, {Ravindranath}, {Scholz}, {Abraham}, {Artigau}, {Brada{\v{c}}}, {Goudfrooij}, {Hutchings}, {Iyer}, {Jayawardhana}, {LaMassa}, {Martis}, {Meyer}, {Morishita}, {Mowla}, {Muzzin}, {Noirot}, {Pacifici}, {Rowlands}, {Sarrouh}, {Sawicki}, {Taylor}, {Volk}, \& {Zabl}}]{willott+2022}
{Willott}, C.~J., {Doyon}, R., {Albert}, L., {et~al.} 2022, \bibinfo{title}{{The Near-infrared Imager and Slitless Spectrograph for the James Webb Space Telescope. II. Wide Field Slitless Spectroscopy},} \pasp, 134, 025002, \dodoi{10.1088/1538-3873/ac5158}

\bibitem[{R.~A. {Windhorst} {et~al.}(2023){Windhorst}, {Cohen}, {Jansen}, {Summers}, {Tompkins}, {Conselice}, {Driver}, {Yan}, {Coe}, {Frye}, {Grogin}, {Koekemoer}, {Marshall}, {O'Brien}, {Pirzkal}, {Robotham}, {Ryan}, {Willmer}, {Carleton}, {Diego}, {Keel}, {Porto}, {Redshaw}, {Scheller}, {Wilkins}, {Willner}, {Zitrin}, {Adams}, {Austin}, {Arendt}, {Beacom}, {Bhatawdekar}, {Bradley}, {Broadhurst}, {Cheng}, {Civano}, {Dai}, {Dole}, {D'Silva}, {Duncan}, {Fazio}, {Ferrami}, {Ferreira}, {Finkelstein}, {Furtak}, {Gim}, {Griffiths}, {Hammel}, {Harrington}, {Hathi}, {Holwerda}, {Honor}, {Huang}, {Hyun}, {Im}, {Joshi}, {Kamieneski}, {Kelly}, {Larson}, {Li}, {Lim}, {Ma}, {Maksym}, {Manzoni}, {Meena}, {Milam}, {Nonino}, {Pascale}, {Petric}, {Pierel}, {Polletta}, {R{\"o}ttgering}, {Rutkowski}, {Smail}, {Straughn}, {Strolger}, {Swirbul}, {Trussler}, {Wang}, {Welch}, {B. Wyithe}, {Yun}, {Zackrisson}, {Zhang}, \& {Zhao}}]{windhorst+2023}
{Windhorst}, R.~A., {Cohen}, S.~H., {Jansen}, R.~A., {et~al.} 2023, \bibinfo{title}{{JWST PEARLS. Prime Extragalactic Areas for Reionization and Lensing Science: Project Overview and First Results},} \aj, 165, 13, \dodoi{10.3847/1538-3881/aca163}

\bibitem[{S.~E. {Zepf} {et~al.}(1995){Zepf}, {Ashman}, \& {Geisler}}]{ashman_zepf1995}
{Zepf}, S.~E., {Ashman}, K.~M., \& {Geisler}, D. 1995, \bibinfo{title}{{Constraints on the Formation History of the Elliptical Galaxy NGC 3923 from the Colors of Its Globular Clusters},} \apj, 443, 570, \dodoi{10.1086/175549}

\end{thebibliography}
\bibliographystyle{aasjournalv7}



\end{document}